\documentclass[fleqn,usenatbib]{mnras}

\usepackage{newtxtext,newtxmath}

\usepackage[T1]{fontenc}
\usepackage{ae,aecompl}

\usepackage{amsfonts}
\usepackage{amsmath}
\usepackage{amssymb}
\usepackage[ngerman, british]{babel}
\usepackage{booktabs}
\usepackage{xcolor}
\usepackage{enumitem}
\usepackage{graphicx}
\usepackage[utf8]{inputenc}
\usepackage[normalem]{ulem}

\newcommand{\mum}{\,\mu\hbox{m}}

\definecolor{lgreen}{rgb}{0.0, 0.7, 0.0}

\title[Tracing planetesimal belts]{Dust Spreading in Debris Discs: Do Small Grains Cling on to Their Birth Environment?}

\author[N. Pawellek et al.]{
Nicole Pawellek$^{1,2},$\thanks{E-mail: pawellek@mpia.de}
Attila Mo\'or$^{2}$,
Ilaria Pascucci$^{1,3,4}$,
and
Alexander V. Krivov$^5$
\\
$^{1}$ Max-Planck-Institut für Astronomie, Königstuhl 17, 69117 Heidelberg, Germany\\
$^{2}$ Konkoly Observatory, Research Centre for Astronomy and Earth Sciences, Hungarian Academy of Sciences, \\ Konkoly-Thege Mikl\'os \'ut 15-17, H-1121 Budapest, Hungary\\
$^{3}$ Lunar and Planetary Laboratory, The University of Arizona, Tucson, AZ 85721, USA\\
$^{4}$ Earths in Other Solar Systems Team, NASA Nexus for Exoplanet System Science\\
$^{5}$ Astrophysikalisches Institut und Universitätssternwarte, Friedrich-Schiller-Universität Jena, Schillergässchen 3, 07745 Jena, Germany\\
}

\date{Accepted XXX. Received YYY; in original form ZZZ}

\pubyear{2018}

\begin{document}
\label{firstpage}
\pagerange{\pageref{firstpage}--\pageref{lastpage}}
\maketitle

\begin{abstract}
Debris discs are dusty belts of planetesimals around main-sequence stars, similar to the asteroid and Kuiper belts in our solar system. The planetesimals cannot be observed directly, yet they produce detectable dust in mutual collisions. Observing  the dust, we can try to infer properties of invisible planetesimals. Here we address the question of what is the best way to measure the location of outer planetesimal belts that encompass extrasolar planetary systems. A standard method is using resolved images at mm-wavelengths, which reveal dust grains with sizes comparable to the observational wavelength. Smaller grains seen in the infrared (IR) are subject to several non-gravitational forces that drag them away from their birth rings, and so may not closely trace the parent bodies. In this study, we examine whether imaging of debris discs at shorter wavelengths might enable determining the spatial location of the exo-Kuiper belts with sufficient accuracy. We find that around M-type stars the dust best visible in the mid-IR is efficiently displaced inward from their birth location by stellar winds, causing the discs to look more compact in mid-IR images than they actually are. 
However, around earlier-type stars where the majority of debris discs is found, discs are still the brightest at the birth ring location in the mid-IR regime. Thus, sensitive IR facilities with good angular resolution, such as MIRI on JWST, will enable tracing exo-Kuiper belts in nearby debris disc systems.
\end{abstract}




\section{Introduction}

Debris discs are dusty belts of comets and asteroids around stars, left over in planetary systems as a by-product of planet formation. These small bodies, or planetesimals, remain invisible to our telescopes. However, they produce detectable dust grains in mutual collisions and other destructive processes. Thus the properties of these planetesimals can only be inferred from observations of their dust \citep{wyatt-2008, krivov-2010, hughes-et-al-2018, wyatt-et-al-2018}.

Observations at different wavelengths trace dust particles of different sizes,
with large grains being best probed by observations at long wavelengths.
The main reason is that the emission efficiency of dust particles drops at wavelengths longer than their sizes \citep[e.g., see Figs. in][]{draine-lee-1984, david-pegourie-1995}. Consequently, long-wavelength observations show large grains only, whereas short-wavelength emission reveals a mixture of large and small grains, rendering the interpretation more difficult. 
However, large grains are generally expected to be less abundant than smaller ones. Particles with sizes above centimetres remain invisible at any wavelength, simply because the cross section they carry is too small for any reasonable slope of the dust grain size distribution. This obviously implies to objects of interest, planetesimals, which raises the question of how we can trace them by observing the emission of dust. 

\vspace*{1cm}

To answer this question, it is important to understand the behaviour of dust grains of different sizes. The orbits of the dust particles are altered by numerous mechanisms such as radiation pressure, collisions and drag forces \citep[e.g., ][]{burns-et-al-1979, wyatt-et-al-1999, krivov-et-al-2006, wyatt-et-al-2007}, with small grains being affected more strongly compared to their larger siblings. As a result, large grains should be located in close proximity of their parent bodies. This would suggest observing debris discs at the longest wavelengths possible, typically millimetre wavelengths, to trace the planetesimals. An intercessor for this assumption are near-infrared scattered light observations which show a more extended geometry of the discs compared to millimetre measurements and thus, make the tracing of planetesimal belts difficult \citep[e.g., $\beta$~Pic, HD~15115, HD~32297:][]{ballering-et-al-2016, engler-et-al-2019, macgregor-et-al-2018}.

However, for a given aperture size of an instrument, e.g., for the Mid-Infrared Instrument (MIRI) on the upcoming James Webb Space Telescope \citep[JWST,][]{greenhouse-2016}, or an interferometric baseline length, e.g., for the Atacama Large Millimeter/submillimeter Array (ALMA), the achievable angular resolution is proportional to the wavelength. This, together with the fact that large facilities, such as JWST or the Extremely Large Telescope \citep[ELT,][]{padovani-2018} operating at shorter (mid- or even near-infrared) wavelengths will become available in the near future, reinforces the question of what the ``best'' wavelength is to trace planetesimal belts. 
Could high-resolution observations with a large telescope enable to probe the location of planetesimals at wavelengths shorter than millimetre?
This might particularly be interesting for the upcoming JWST operating at mid-infrared wavelengths. We will be able to spatially resolve debris discs with JWST and thus, it might be possible to trace the planetesimal belts located close-in to the stars at least as accurately as it can currently be done for outer belts with ALMA \citep[e.g., ][]{moor-et-al-2013, lieman-sifry-et-al-2016, matra-et-al-2018}. 

In this paper, we attempt to clarify whether observations at millimetre wavelengths are really necessary to pinpoint planetesimal belts. The paper is structured as follows. In section~\ref{sec:processes} we explain the physical processes altering the particle orbits. Section~\ref{sec:modelling} will describe our modelling approach, which is to simulate the spatial distribution of dust grains in relation to their birth ring for three fiducial main-sequence host stars of different spectral types.
The results are presented in section~\ref{sec:results} and the conclusions are drawn in section~\ref{sec:conclusions}.

\section{Alteration of particle orbits}
\label{sec:processes}

Dust particles observed in debris discs are subject to a large array of forces and effects. In addition to stellar gravity, these include stellar radiation pressure, mutual collisions, drag forces, perturbations of known or alleged planets, and others. We start with a brief description of radiation pressure (and its corpuscular analogue, stellar wind pressure) and collisions, which represent the dominant mechanisms in the majority of detectable debris discs \citep{wyatt-2005}.

\subsection{Radiation pressure}

Stellar radiation pressure alters the orbital motion of dust grains in planetary systems by transferring momentum from stellar photons to the particles. The ratio of the radiation pressure to gravity force is given by \citep{burns-et-al-1979}
\begin{equation}
	\beta \equiv \frac{\left|\vec{F}_\text{rad}\right|}{\left|\vec{F}_\text{G}\right|} = \frac{3 L}{16\pi G c M}\times \frac{Q_\text{pr}}{\varrho s},
    \label{eq:beta}
\end{equation} 
where $L$ and $M$ are the stellar luminosity and mass, $G$ the gravitational constant, $c$ the speed of light, $s$ the radius of a (spherical) particle, $\varrho$ the particle's bulk density, and $Q_\text{pr}$ the radiation pressure efficiency averaged over the stellar spectrum.

\begin{figure}
\centering
	\includegraphics[width=0.35\textwidth, angle=-90]{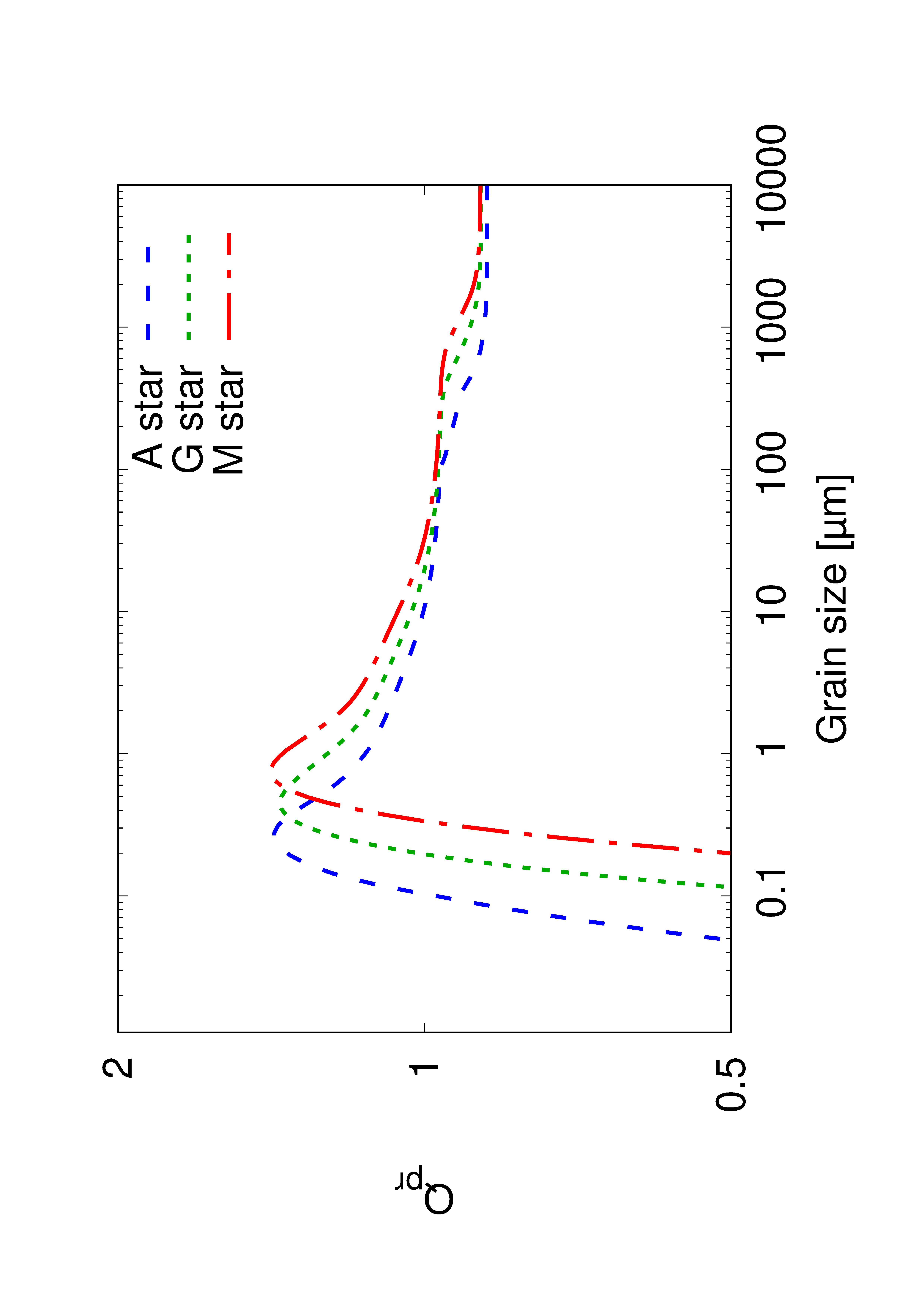}
    \caption{Radiation pressure efficiency as a function of grain size. 
    }
     \label{fig:Qpr}
\end{figure}

The larger $\beta$ is, the stronger the dust particle is affected by stellar radiation pressure.
While independent of the distance to the star, $\beta$ is inversely proportional to the particle size, $s$. Furthermore, $\beta$ depends on the radiation pressure efficiency, $Q_\text{pr}$, which is also dependent on the particle size. Fig.~\ref{fig:Qpr} shows the relation between $Q_\text{pr}$ and the grain size, assuming 
astronomical silicate \citep{draine-2003} with a bulk density of $3.3~$g/cm$^3$ as dust composition \footnote{Using different dust compositions has only a minor effect on the results of this analysis. Therefore, we use astrosilicate following prior studies.}.
We calculated $Q_\text{pr}$ with the code of \cite{bohren-huffman-1983}, which is based on the Mie theory.
For small grains $Q_\text{pr}$ shows a steep increase while for large grains it stays nearly constant. 
Radiation pressure drives dust grains released from the parent bodies into larger and more eccentric orbits than those of the parent planetesimals. As soon as $\beta$ is known, one can calculate the orbital elements of the dust grains upon release by using the formulae given, e.g., in \citet{burns-et-al-1979,wyatt-et-al-1999,krivov-et-al-2006}.

\subsection{Poynting-Robertson drag and stellar wind}

Apart from the direct radiation pressure outlined above, which acts in radial direction, a grain is subject to a tangential Poynting-Robertson (P-R) force. It causes a grain to lose its energy and angular momentum and thus to slowly spiral from its birth location towards the star.

The significance of P-R drag for debris discs was analysed both with help of observations \citep[e.g.,][]{mennesson-et-al-2014, ertel-et-al-2018} and theoretically \citep[e.g.,][]{wyatt-2005, kennedy-piette-2015}.

\cite{kennedy-piette-2015} use the normal geometric optical depth to illustrate the importance of P-R drag:
\begin{equation}
    \tau(r) = \frac{\tau(r_0)}{1+4\eta_0\left(1-\sqrt{r/r_0}\right)},
    \label{tau}
\end{equation}
with
\begin{equation}
    \eta_0 = \frac{5000}{k \beta} \; \tau(r_0) \; \sqrt{\left(\frac{r_0}{a_\text{earth}}\right) \frac{M_\odot}{M}},
    \label{eta0}
\end{equation}
where $r_0$ is the distance at which the parent planetesimal belt is located, $a_\text{earth}= 1$au, and $M_\odot$ is the solar mass.
The factor $k$ depends on the collisional model employed.
It is unity in the analytic model of \cite{wyatt-2005} that assumed equal-sized grains. In addition to this, \cite{kennedy-piette-2015} found a value of 1/7 to better match more accurate numerical models that assumed a distribution of grain sizes. Following \cite{kennedy-piette-2015}, we refer to $k=1$ and $k=1/7$ as  the ``low collisional rate''  and ``high collisional rate'' regimes, respectively.

Like photons radiated by the central stars, stellar wind particles exert pressure on the dust grains. Similarly to radiation pressure, the resulting stellar wind force has two components, a radial one (direct stellar wind pressure) and a tangential one (stellar wind drag, acting qualitatively in the same way as P-R drag). While the former is generally small and can be ignored,  the latter may be important. For late-type stars that possess strong winds, stellar wind drag may exceed P-R drag by several orders of magnitude \citep[e.g.,][]{plavchan-et-al-2005, schueppler-et-al-2015}.

\subsection{Collisions}

Relative velocities between the particles in debris discs are typically high enough for collisions to be erosive or destructive, creating smaller fragments.
Models describe the frequency of particle collisions and the collisional outcome \citep[e.g.,][]{dohnanyi-1969, strubbe-chiang-2006, thebault-wu-2008}.
The frequency of collisions is related to the probability that two grains on Keplerian orbits come close enough to touch \citep{krivov-et-al-2005}. 
The collisional outcome depends on the physics of the impact and the fragmentation process and ranges from elastic collisions to totally inelastic ones, with and without fragmentation. 
The physics of collisions and its implementation in the modelling codes are described in previous studies \citep[e.g.][]{thebault-et-al-2003, grigorieva-et-al-2007, mueller-et-al-2009, loehne-et-al-2017}.

\begin{table*}
\caption{A list of model setups.}
	\begin{tabular}{llrllrrrrrrr}
	$L/L_\odot$	& $M/M_\odot$	& T$_\text{eff}$ [K]& 	$s_\text{blow}$ [$\mu$m] & $\dot{M}$ [$\dot{M_\odot}$] & PR-drag	& $R$ [AU] & $\Delta R$ [AU]	& $\langle e \rangle$ & $M_\text{disc}$ [$M_\text{earth}$] &$\lambda$ [$\mum$]& Figures\\
    \toprule
    0.01	& 0.30	& 3500	& \ldots  & \ldots  & off & 30	 & 3	  & 0.1	  & 1	 & 23, 100, 1000     & \ref{fig:ACE_Mstar}, \ref{fig:ACE_Mstar_Profile}\\
   			&       &  		& \ldots  & \ldots  & off & 100	 & 10	  & 0.1	  & 30   & 23, 100, 1000     & \ref{fig:ACE_Mstar}, \ref{fig:ACE_Mstar_Profile}\\
            & 		&  		& \ldots  & \ldots  & off & 200	 & 20	  & 0.1	  &200   & 23, 100, 1000     & \ref{fig:ACE_Mstar}, \ref{fig:ACE_Mstar_Profile}\\
            & 	    &      	& \ldots  & 50      & off & 30	 & 3	  & 0.1	  & 1	 & 23, 100, 1000     & \ref{fig:ACE_Mstar_Wind}, \ref{fig:ACE_Mstar_Wind_Profile}\\
   			& 		&  		& \ldots  & 50      & off & 100	 & 10	  & 0.1	  & 30   & 23, 100, 1000     & \ref{fig:ACE_Mstar_Wind}, \ref{fig:ACE_Mstar_Wind_Profile}\\
            & 		&  		& \ldots  & 50      & off & 200	 & 20	  & 0.1	  &200   & 23, 100, 1000     & \ref{fig:ACE_Mstar_Wind}, \ref{fig:ACE_Mstar_Wind_Profile}\\
     \midrule
    1.00	&	1.0  & 5800	 & 0.50  & \ldots  & off & 30     & 3   & 0.1  & 1    & 23, 100, 1000     &\ref{fig:ACE_Gstar}, \ref{fig:ACE_Gstar_Profile}\\
            &        &       & 0.50  & \ldots  & off & 100    & 10  & 0.1  & 30   & 23, 100, 1000     &\ref{fig:ACE_Gstar}, \ref{fig:ACE_Gstar_Profile}\\
            &        &       & 0.50  & \ldots  & off & 200    & 20  & 0.1  & 200  & 23, 100, 1000     &\ref{fig:ACE_Gstar}, \ref{fig:ACE_Gstar_Profile}\\
            &        &       & 0.50  & \ldots  & on  & 100    & 10  & 0.1  & 0.1, 1, 10, 30 & 23, 100, 1000 & \ref{fig:PR-drag}, \ref{fig:PR-drag_thermalemission}, \ref{fig:PR-drag_profile}\\  
     		&		 &  	 & 0.50  & \ldots  & off & 100	 & 20  & 0.1  & 30   & 23, 100, 1000     &\ref{fig:ACE_Gstar_Width}, \ref{fig:ACE_Gstar_Width_Profile}\\
            &		 &  	 & 0.50  & \ldots  & off & 100	 & 10  & 0.2  & 30   & 23, 100, 1000     &\ref{fig:ACE_Gstar_Excitation}, \ref{fig:ACE_Gstar_Excitation_Profile}\\
   \midrule
    30.0	& 2.45  & 10000 & 4.44   & \ldots & off & 30   & 3   & 0.1  & 1    & 23, 100, 1000     &\ref{fig:ACE_Astar}, \ref{fig:ACE_Astar_Profile}\\
            &       &       & 4.44   & \ldots & off & 100  & 10  & 0.1  & 30   & 23, 100, 1000     &\ref{fig:ACE_Astar}, \ref{fig:ACE_Astar_Profile}\\
            &       &       & 4.44   & \ldots & off & 200  & 20  & 0.1  & 200  & 23, 100, 1000     &\ref{fig:ACE_Astar}, \ref{fig:ACE_Astar_Profile}\\
 
    \bottomrule
	\end{tabular}
    \label{tab:StellarProperties}
    \newline
    
    Notes: The blow out grain size, $s_\text{blow}$, is calculated for compact spherical particles\\
    made of pure astronomical silicate with a bulk density of $3.3$g/cm$^3$.
\end{table*}

\subsection{Other effects}

Planetary systems may also include additional components, such as the gas environment or perturbing planets. 
For instance, CO gas has been detected in a handful of young debris discs around early-type stars \citep[e.g., ][]{moor-et-al-2017, kral-et-al-2018}. 
However, detectable amount of CO has only been found in some of the debris discs, and only those around young early-type stars. Thus, in the vast majority of debris discs it is unlikely that secondary gas would affect the dust.
Therefore, we do not include gas in this study. In contrast, dynamical effects of planets may be important. Nevertheless, there are only a few systems found to possess both planets and debris discs \cite[e.g.,][]{lagrange-et-al-2010, marois-et-al-2010, moromartin-et-al-2010, rameau-et-al-2013}, and in most of these cases the planets are located far away from the cold dust disc \cite[e.g.,][]{santos-et-al-2000, butler-et-al-2003, bonfils-et-al-2005}. Such planets should not have a significant effect on the dust distributions investigated here. Of course, additional, as yet undiscovered, planets may be present \citep[e.g.,][]{marino-et-al-2018b} and they may affect the discs in various ways\citep[e.g.,][]{ertel-et-al-2012, bonsor-et-al-2018, wyatt-et-al-2018}. However, the masses and orbits of these planets are unknown. To keep the analysis simple and to avoid dealing with an unmanageable number of free parameters, we do not consider planetary perturbations in our analysis either.

\section{Modelling approach}
\label{sec:modelling}

In this section, we describe a representative grid of central stars, their debris discs, and the modeling procedure.

\subsection{Host stars}

Although debris discs were found around stars of various evolutionary stages, including sub-giant stars and white dwarfs \citep{matthews-et-al-2014}, the majority of them are hosted by main-sequence stars. 
Therefore, in the following we adopt main-sequence host stars with three different spectral types to cover a broad range in stellar properties: an M3-type, a G2-type, and an A1-type star. 
We use the photospheric models given by \cite{hauschildt-et-al-1999} and assume typical values for the stellar luminosities, temperatures and masses (see Table~\ref{tab:StellarProperties}) to calculate the radiation pressure and dust temperature.

M-stars are known to possess strong stellar winds and thus, it is of interest to include the wind into the M-star models.
Following \citet{schueppler-et-al-2015}, we adopt a wind speed of 400~km/s and a mass loss rate of 50 times the solar one, which amounts to $2\times10^{-14}M_\odot/\text{yr}$. 
For comparison, however, we also generate models neglecting the wind.

\subsection{Planetesimal belts}

We generate parent belts at three different distances for each host star: 30, 100 and 200~AU assuming a flat radial distribution with a radial distribution index of 0.0, a free initial eccentricity of 0.1 and a relative width of $\pm10$\%. The narrowness of these belts is justified since the majority of debris discs are found to be confined into narrow rings \citep[e.g.,][]{kennedy-wyatt-2010}. 
However, in recent studies discs were also found to possess extended planetesimal belts \citep[e.g.,][]{matra-et-al-2018}, which makes the additional investigation of a broad ring model a logical consequence.
To get comparable spatial densities we use the initial disc masses given in \cite{krivov-et-al-2008}:
for discs at 30~AU a mass of 1~M$_\text{earth}$ in the bodies smaller than 50 km is assumed while for 100~AU the mass is 30~M$_\text{earth}$ and for 200~AU it is 200~M$_\text{earth}$.
Several groups have investigated whether there is a relation between the disc radius and the stellar luminosity \citep[e.g.,][]{eiroa-et-al-2013, pawellek-et-al-2014, matra-et-al-2018}. As there is no consensus on the existence of such a relation, we use the same distance of the planetesimal belt to the host star for all spectral types.
We further assume an average disc eccentricity of 0.1, but explore a more excited disc with a value of 0.2 as well \citep{thebault-wu-2008}.

\subsection{Modelling with ACE}

For our analysis we use the modelling code "Analysis of Collisional Evolution" \citep[ACE, see][and references therein]{loehne-et-al-2017} which allows a full featured collisional modelling of debris discs. 
It produces spatial and size distributions of circumstellar material, sized between dwarf planets and dust particles, and evolves them over giga-year time spans.
To do that the code solves the Boltzmann-Smoluchowski kinetic equation over a four-dimensional grid of phase space variables. This approach is different from N-body simulations or hydrodynamical codes. 
ACE was previously used for numerous studies of debris discs \citep[e.g.,][]{krivov-et-al-2005, krivov-et-al-2006, loehne-et-al-2008, loehne-et-al-2012, krivov-et-al-2013}.

ACE allows one to cover a rich set of physical effects acting on the debris material. Of these, we only include stellar gravity, direct radiation pressure, and realistic collisional outcomes, such as sticking, rebounding, cratering and disruption, and collisional damping.
Besides, we add stellar wind drag and P-R drag in some of the cases. 

Unless otherwise stated, the disc evolution was followed for 100~Myr. This choice is not essential, as a steady state is typically reached on shorter timescales (except for the runs with the stellar wind and P-R drag, see remarks in Sect.~4.1.2 and 4.2.2).

\subsection{Thermal emission calculations}

For each of the ACE simulations, where we assume pure astronomical silicate \citep{draine-2003} as dust material, thermal emission maps are generated.
We calculate the absorption efficiencies for all dust particles with the help of the Mie theory \citep[e.g.,][]{bohren-huffman-1983} and infer the grain temperatures using this parameter and the particles' distance to the star.
From the grain temperature we infer the flux density of the optically thin dust emission and add it up to an emission map. 
We also generate radial profiles of the surface brightness. 

All these calculations are done for three representative wavelengths, 23~$\mum$, 100~$\mum$, and 1000~$\mum$.
The wavelengths were chosen because of their relevance to debris disc studies with past and current facilities, i.e. $23\mum$ (close to $24\mum$ of Spitzer/MIPS, e.g. \citealt{rieke-et-al-2004}), $100\mum$ (Herschel/PACS, e.g. \citealt{poglitsch-et-al-2010}) and $1000\mum$ (ALMA), 
as well as to explore the potential of JWST/MIRI \citep[$23\mum$, ][]{rieke-et-al-2015} to characterise debris discs.

\section{Results}
\label{sec:results}

In this section, we present and discuss the simulation results from all setups summarised in Table~\ref{tab:StellarProperties}.
The figures illustrating the results (e.g., Figs.~\ref{fig:ACE_Mstar}--\ref{fig:ACE_Astar_Profile}) are all organised in a similar way. 
The columns from left to right always correspond to the different disc radii of 30, 100 and 200~AU, while panels from top to bottom present three wavelengths, 23~$\mum$, 100~$\mum$, and 1000~$\mum$. We normalise the surface brightness to its maximum to keep the same colour scale for each image. We define the radius of a debris disc seen in thermal emission as the location of the maximum surface brightness. The extent of a disc is defined as a distance between the inner and outer positions where the surface brightness drops to $1/e\approx37\%$ of its maximum value.

\subsection{Late-type stars}

\subsubsection{Without stellar wind}

\begin{figure*}
\centering
	\includegraphics[width=0.8\textwidth]{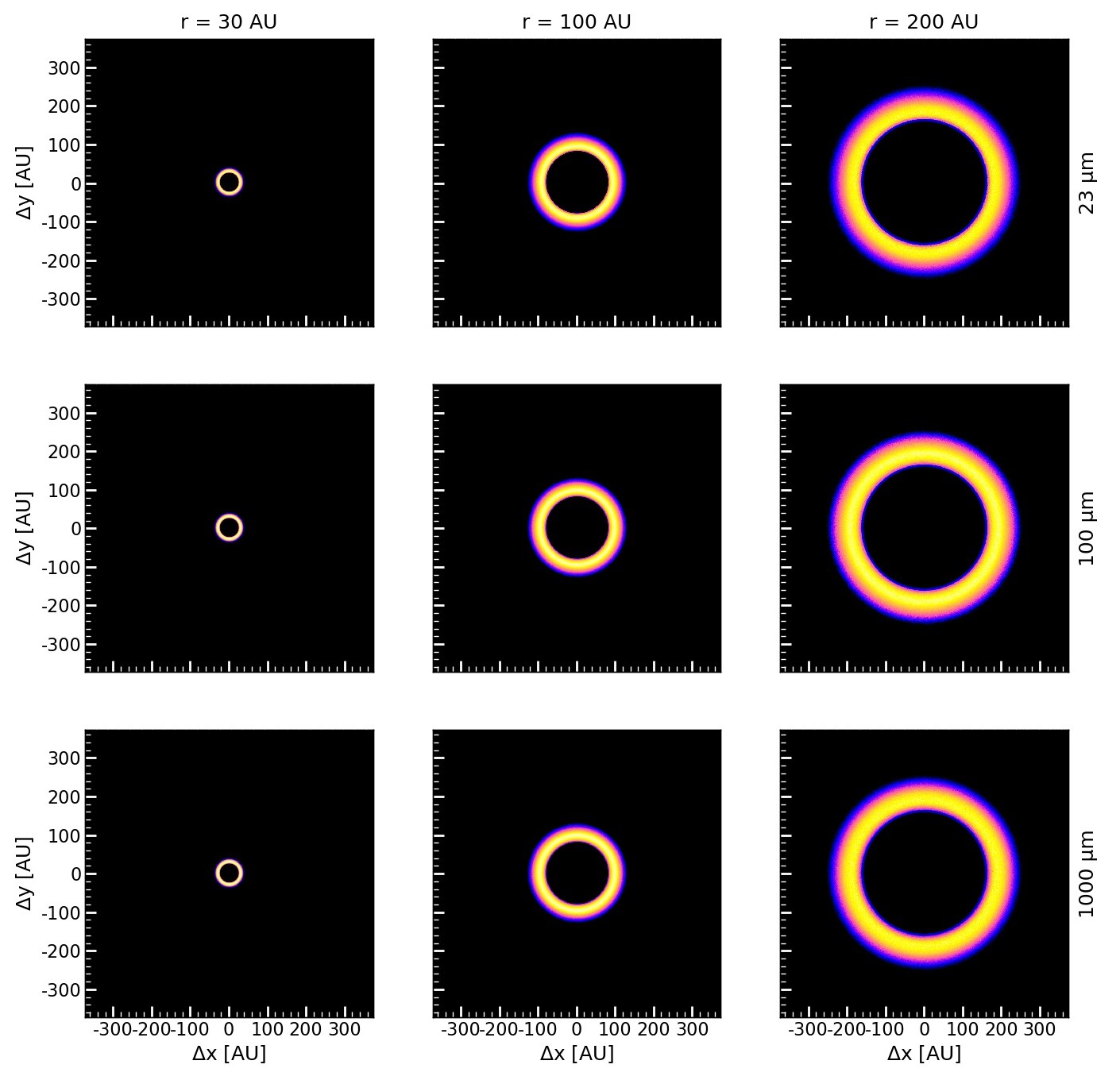}
    \caption{Thermal emission images for an M-type star with $L/L_\odot=0.01$, without stellar wind. From top to bottom: observational wavelengths of 23, 100 and 1000$\mum$. From left to right: planetesimal disc radii of 30, 100 and 200~AU. Since the central star has a low luminosity, all of these dust rings are cold and thus emit very weakly at 23$\mum$. }
     \label{fig:ACE_Mstar}
\end{figure*}

\begin{figure*}
\centering
	\includegraphics[width=\textwidth]{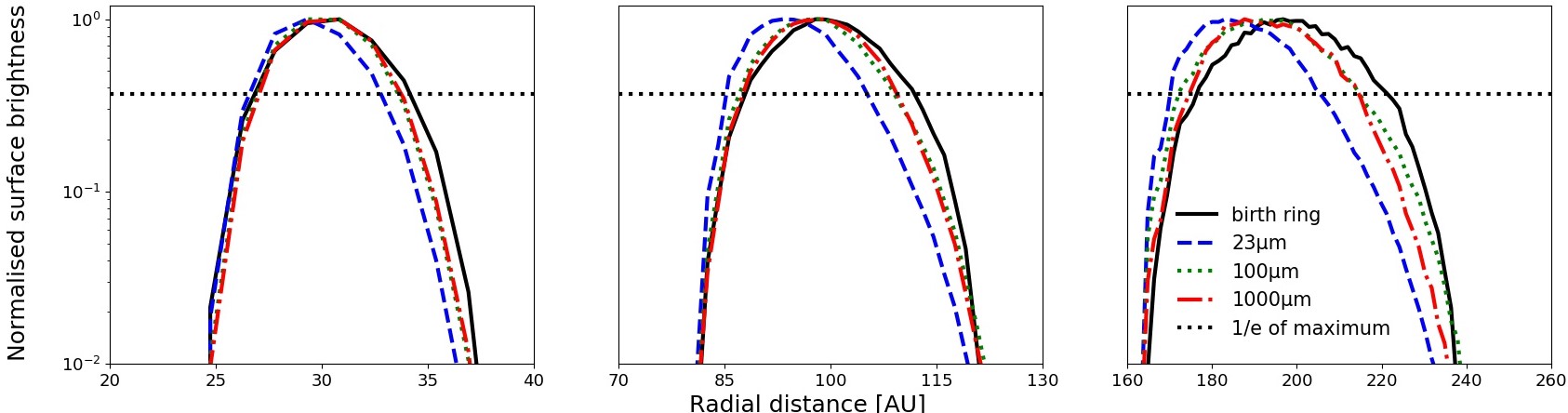}
    \caption{Surface brightness profile for an M-type star with $L/L_\odot=0.01$ and without stellar wind. From left to right: planetesimal disc radii of 30, 100 and 200~AU. The black solid line shows the birth ring, while the profile is presented with a blue dashed line at 23~$\mum$, a green dotted line at 100~$\mum$ and the red dash-dotted line at 1000~$\mum$. The black dotted line shows the $1/e$ level of the maximum.
    }
     \label{fig:ACE_Mstar_Profile}
\end{figure*}

The thermal emission images and radial profiles are depicted in Figs.~\ref{fig:ACE_Mstar} and \ref{fig:ACE_Mstar_Profile}. 
We assume a late-type star with $L/L_\odot = 0.01$.
For low luminosity stars the parameter $\beta$ stays well below 0.5 for all grain sizes \citep[see Fig.~6.6 in][]{pawellek-2017}. This means that no particle is expelled from the system due to radiation pressure and furthermore the particle orbits are not significantly altered. Thus, we expect the dust grains to be located close to their parent belt and see our expectations confirmed by looking at the figures mentioned above. There is no disc broadening at any observational wavelength used in this study. However, there are slight changes in the radial profiles compared to the birth ring due to the numerical modelling effects (Fig.~\ref{fig:ACE_Mstar_Profile}).
Nevertheless, the maximum of the surface brightness stays close to the birth ring maximum with a deviation of $\sim$5\%. 
Since the dust particles stay close to the parent belt we do not see brightness variations at the different wavelengths chosen.

\subsubsection{With stellar wind}

\begin{figure*}
\centering
	\includegraphics[width=0.80\textwidth]{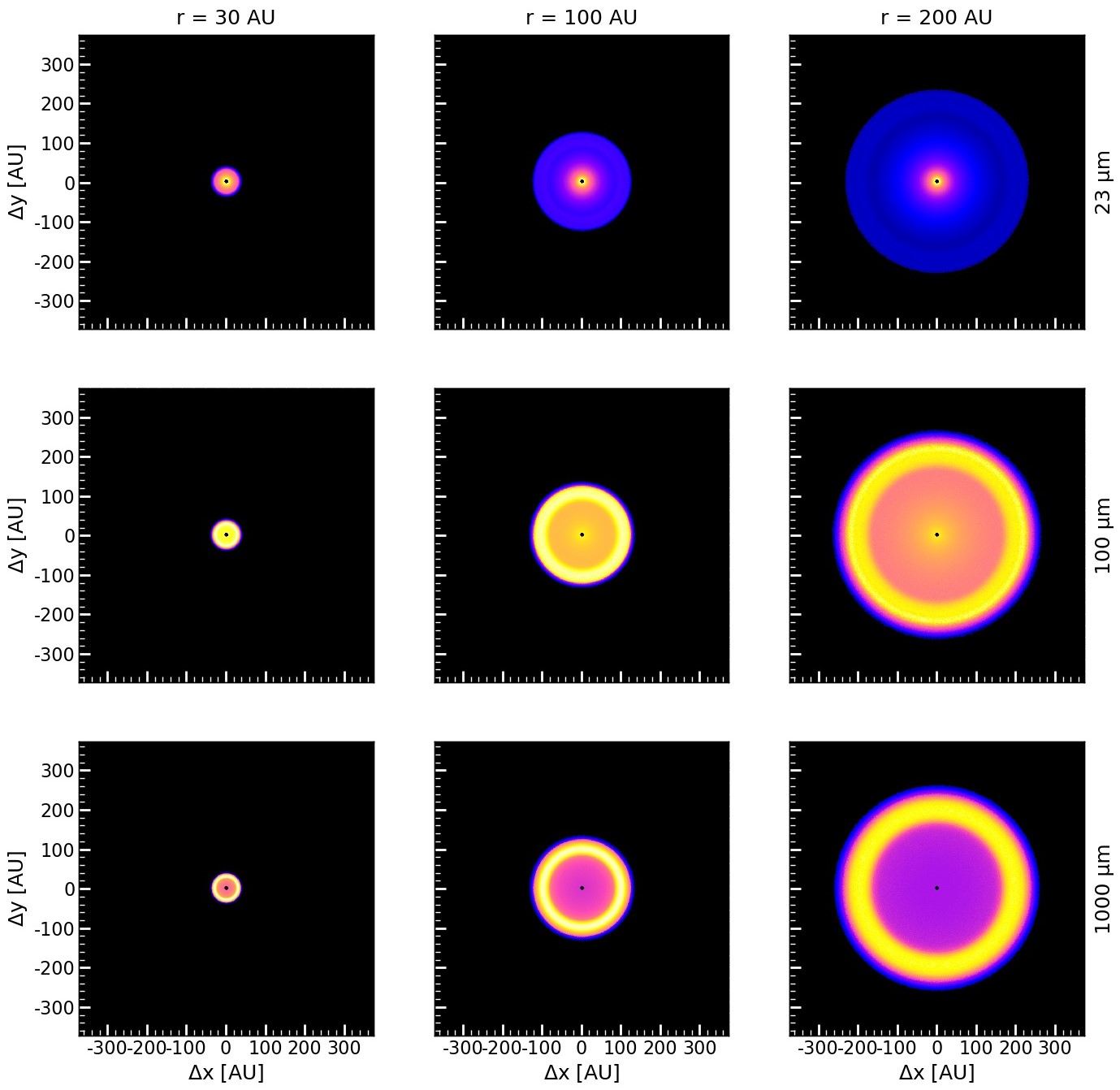}
    \caption{Same as Fig.~\ref{fig:ACE_Mstar}, but including stellar wind.
    }
     \label{fig:ACE_Mstar_Wind}
\end{figure*}

\begin{figure*}
\centering
	\includegraphics[width=\textwidth]{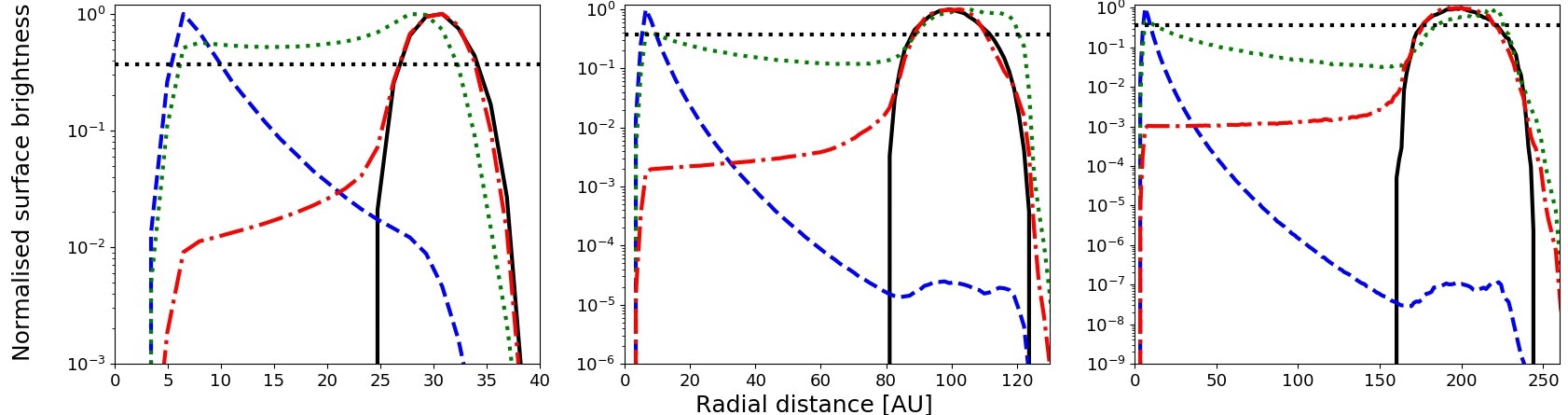}
    \caption{Same as Fig.~\ref{fig:ACE_Mstar_Profile}, but including stellar wind. 
    }
     \label{fig:ACE_Mstar_Wind_Profile}
\end{figure*}

Although stellar radiation pressure is not significant for late-type stars, the stellar wind is expected to be stronger around these stars compared to their earlier-type siblings \citep[e.g.,][]{plavchan-et-al-2005,plavchan-et-al-2009, schueppler-et-al-2014}.
The stellar wind can increase the $\beta$ parameter of the dust particles \citep[see Fig.~1 in ][]{schueppler-et-al-2015} and thus expel the smallest grains from the system. 
Thus, we assume the same late-type star with $L/L_\odot = 0.01$, but add a stellar wind with a velocity of 400~km/s and a mass loss rate of 50 times the solar value.
These values are close to those that were found for AU~Mic in a quiescent state \citep{augereau-beust-2006}.
The resulting thermal emission maps and radial profiles are depicted in Figs.~\ref{fig:ACE_Mstar_Wind} and \ref{fig:ACE_Mstar_Wind_Profile}.


Stellar wind causes the dust particles to drift inwards and to start filling the inner region of the disc. Again, the smaller grains are more affected by this process and thus, we find that the majority of small particles does not stay at the place of the planetesimal belt, but are located close to the star (around $\sim$5~AU) after a 100~Myr time span. As a result, the maximum of the surface brightness is displaced inwards from the birth ring location, which makes the tracing of planetesimals in the mid-infrared impossible. However, for larger grains the surface brightness peak stays at the birth ring position.
An example for this effect might be given in the debris disc around the M-dwarf  WISE-J080822.18644357.3 \citep{flaherty-et-al-2019}. Its SED is showing a strong near-infrared excess with an inferred temperature of $\sim$1000~K. Simultaneously, a sub-mm excess was found for this object.

As a word of caution, we note the results of all the runs involving transport 
would depend on the amount of dust in the disc (i.e., dust mass, fractional luminosity, or any other equivalent quantity). The dustier the disc, the less the displacement of dust inward~-- and thus the stronger the brightness peak \citep[e.g.,][]{wyatt-2005}.

 Another issue associated with the action of stellar wind is that the resulting brightness profiles would depend on the assumed time interval of the disc evolution. If some transport processes are operating, there is no steady-state in the classical sense. Smaller grains drift inwards faster than larger ones. Thus in younger systems the cavities only contain smaller grains, whereas in older systems they should also contain larger ones. As a result, in younger discs one can trace the birth ring at shorter wavelengths than in older ones.

\subsection{Solar-type stars}

\subsubsection{Without P-R drag}

\begin{figure*}
\centering
	\includegraphics[width=0.8\textwidth]{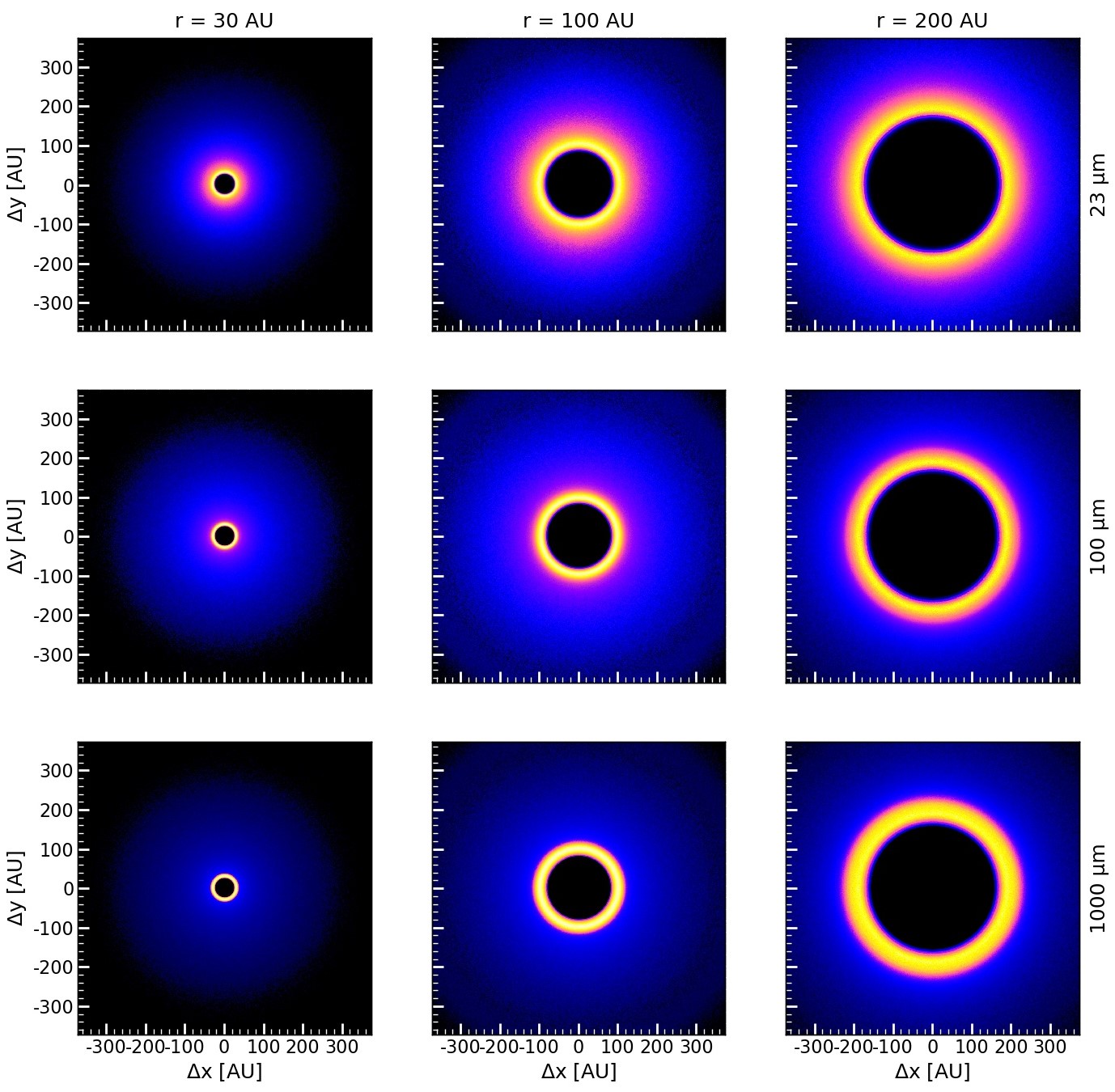}
    \caption{Same as Fig.~\ref{fig:ACE_Mstar}, but for a G-type star with $L/L_\odot=1.0$.}
     \label{fig:ACE_Gstar}
\end{figure*}

\begin{figure*}
\centering
	\includegraphics[width=\textwidth]{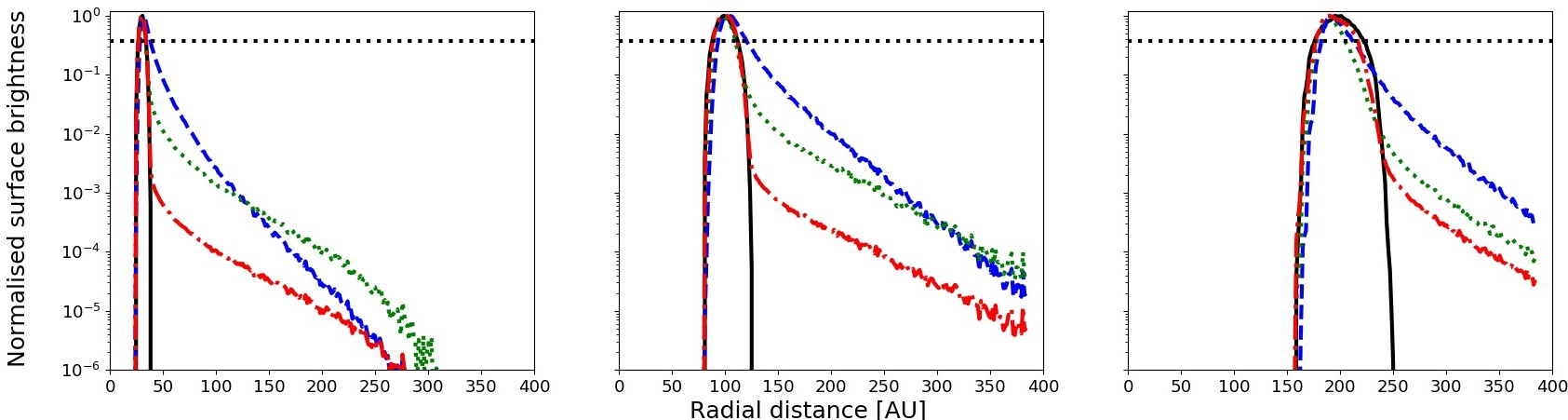}
    \caption{Same as Fig.~\ref{fig:ACE_Mstar_Profile}, but for a G-type star with $L/L_\odot=1.0$.
    }
     \label{fig:ACE_Gstar_Profile}
\end{figure*}

In Fig.~\ref{fig:ACE_Gstar} we show the resulting thermal emission maps for a G-type star with $L/L_\odot=1.0$. The figure panels are the same as in Fig.~\ref{fig:ACE_Mstar}. The corresponding radial profiles are given in Fig.~\ref{fig:ACE_Gstar_Profile}.

In contrast to the M-type star, $\beta$ gets larger than 0.5 for a grain size between 0.07~$\mum$ and 0.5~$\mum$.
Thus, small particles are expected to be on highly eccentric orbits. 
Our assumption is confirmed since all discs show a broadening compared to their birth ring. As defined above we assume the disc extend between the locations of the 1/e value of the surface brightness peak. 
Therefore, the broadening is not prominent for far-inrared and sub-millimetre wavelengths and only subtle ($<5\%$) for the mid-infrared image. Nevertheless, it is visible at lower fractions of the surface brightness maximum at all wavelength bands.
However, the maximum stays at the location of the planetesimal belt for all wavelengths.

\subsubsection{Including P-R drag}

We generated a set of debris disc models with disc radii of 100~AU around the G-type host star and assumed four different disc masses: 0.1, 1, 10 and 30~$M_\text{earth}$. While 30~$M_\text{earth}$ is used as a default case, such a disc is rather massive and might not represent the majority of observable discs. Therefore, we investigate less massive discs which cover a broader range of targets observed. 
In the collisional runs with P-R drag included, we assumed typical ages of the discs of 100~Myr and 1~Gyr. From the modelling results we inferred the optical depths and compared them to the analytical models for discs with high and low collision rates provided by \cite{wyatt-2005} and \cite{kennedy-piette-2015} given by Eqs.~(\ref{tau}) and (\ref{eta0}). In these equations, we assumed dust grains with sizes at the blow-out limit of $\sim0.5\mum$ for G-type stars.

\begin{figure*}
\centering
	\includegraphics[width=0.375\textwidth, angle = -90]{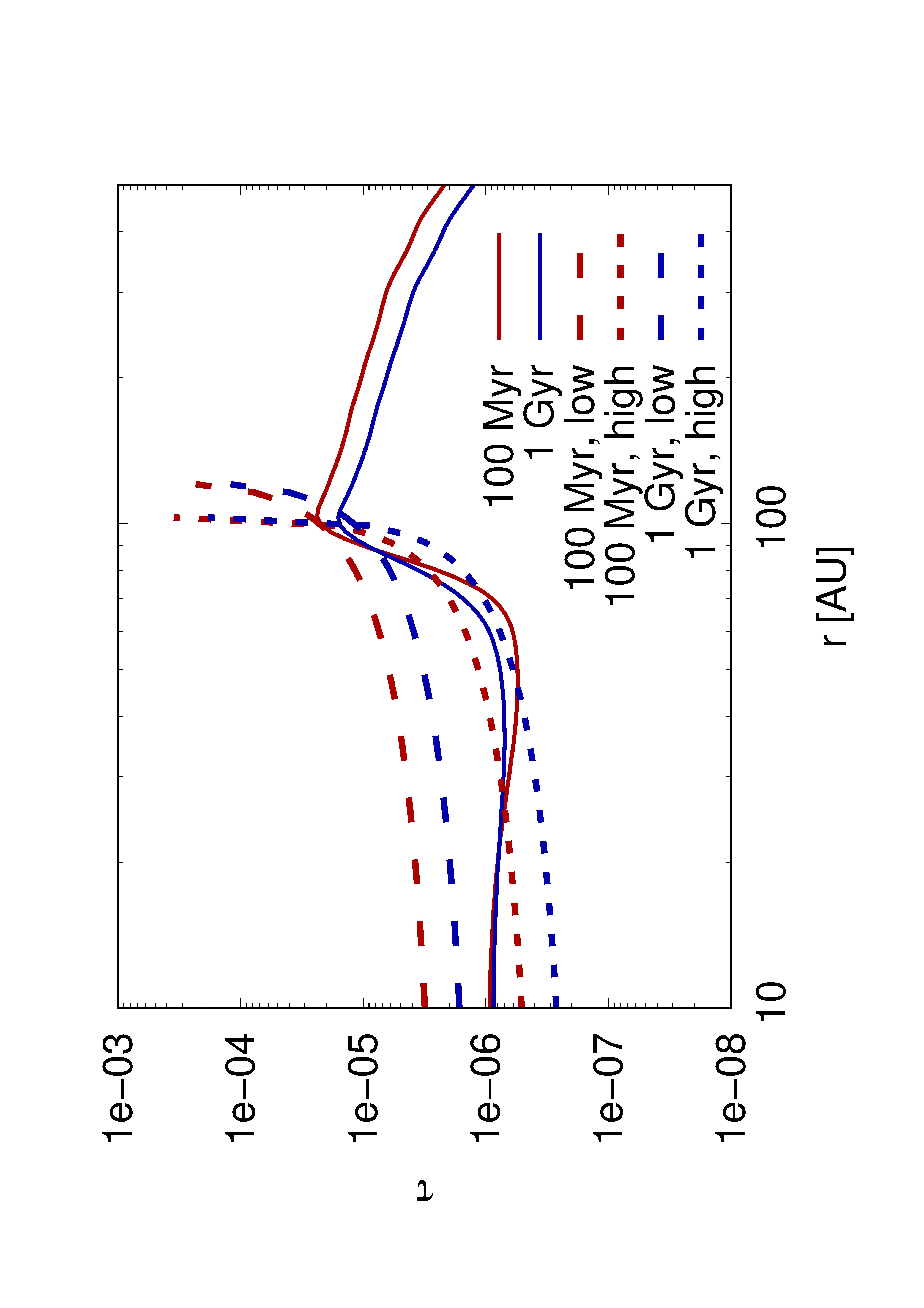}
	\hspace{-1.5cm}
    \includegraphics[width=0.375\textwidth, angle = -90]{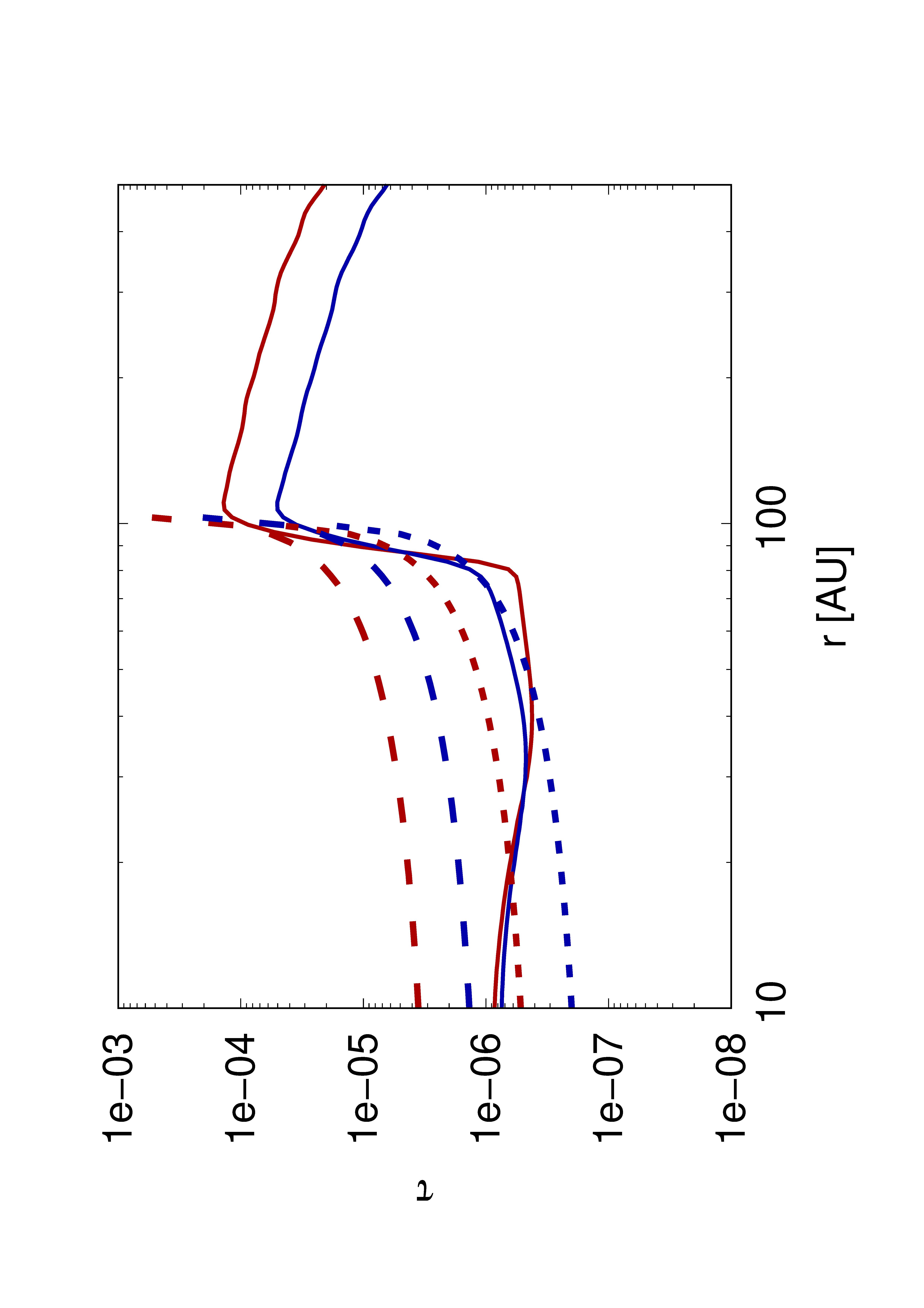}
    \includegraphics[width=0.375\textwidth, angle = -90]{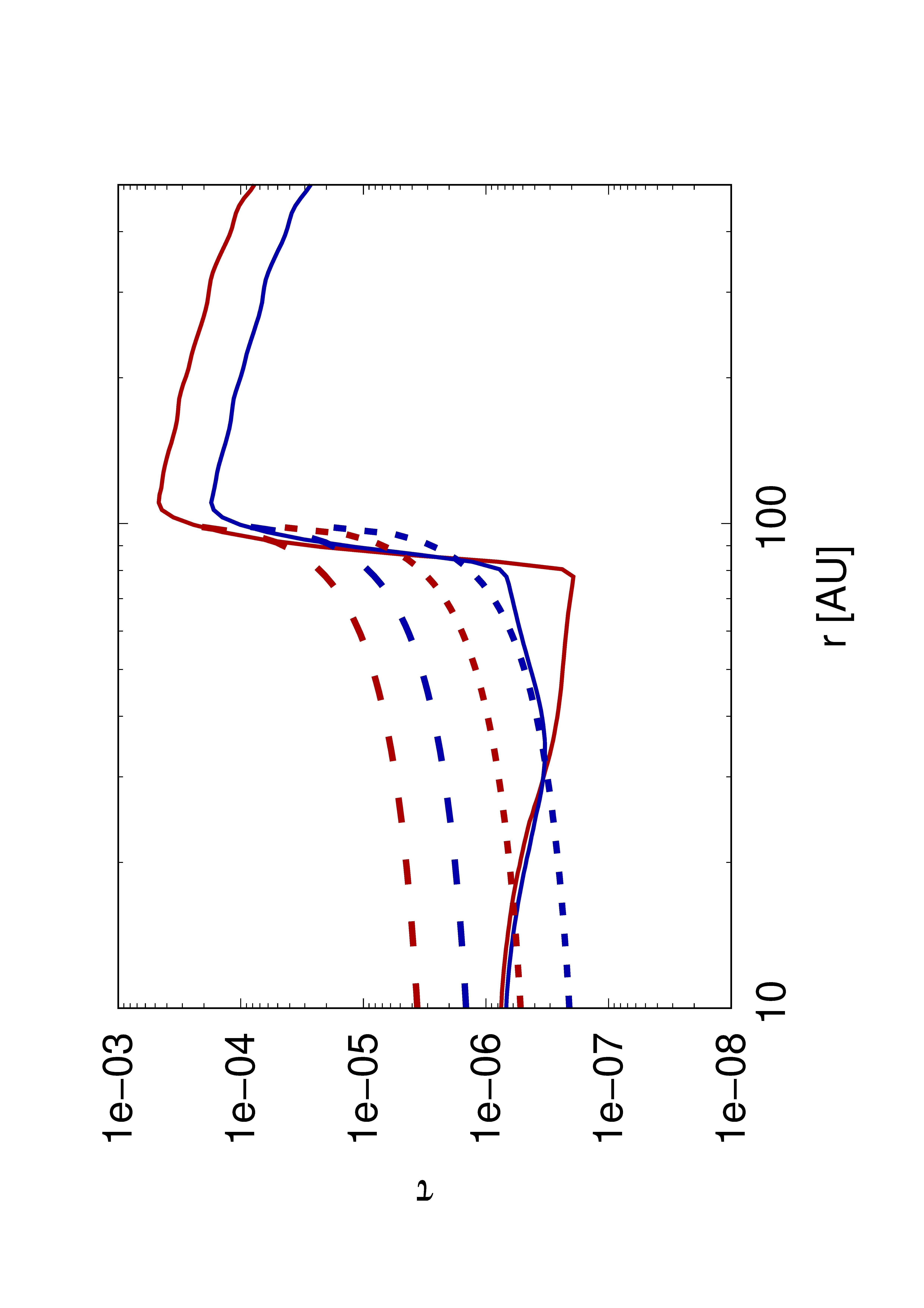}
    \hspace{-1.5cm}
    \includegraphics[width=0.375\textwidth, angle = -90]{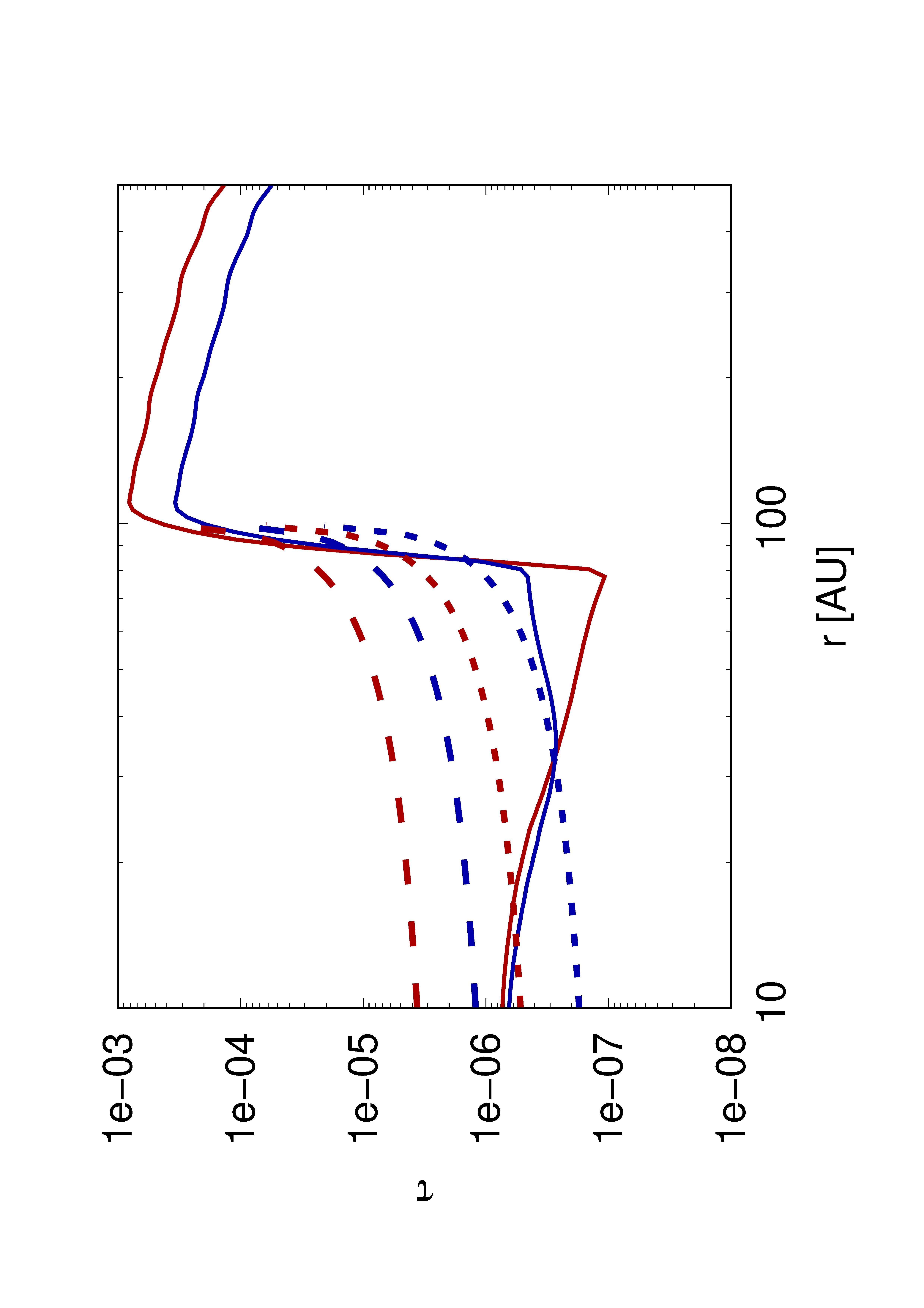}
    \caption{Radial profiles of the optical depth (assuming grains with $\beta=0.5$) for different disc masses around a G-type host star. The planetesimal belt is located at 100~AU. Upper left: 0.1~$M_\text{earth}$; upper right: 1~$M_\text{earth}$; lower left: 10~$M_\text{earth}$; lower right: 30~$M_\text{earth}$. Solid lines: ACE simulations; long-dashed and short-dashed lines: Eqs.~(\ref{tau})--(\ref{eta0}) with $k=1$ and $k=1/7$, respectively. Red: after 100 Myr of collisional evolution, blue: after 1 Gyr.
    }
     \label{fig:PR-drag}
\end{figure*}

The results are presented in Fig.~\ref{fig:PR-drag}. As expected, the optical depth at the planetesimal belt location increases with increasing disc mass. It varies between $10^{-5}$ and $10^{-3}$. The optical depth inside the parent belt drops to $\sim 10^{-6}$, regardless of the disc mass assumed.
The optical depth in the inner region is nearly the same at the ages of 100~Myr and 1~Gyr. The plots also demonstrate that our collisional models are more consistent with high collision rates than low ones. For the disc masses between 1 and 30~$M_\text{earth}$, the difference between the optical depths at the planetesimal belt location and in the inner region is at least one and a half orders of magnitude even for the case of low collision rates. For the disc mass of 0.1~$M_\text{earth}$, the ACE model yields a difference of one order of magnitude, while the analytical model with low collision rates predicts only a drop by a factor of 3. 

\begin{figure*}
\centering
	\includegraphics[width=\textwidth]{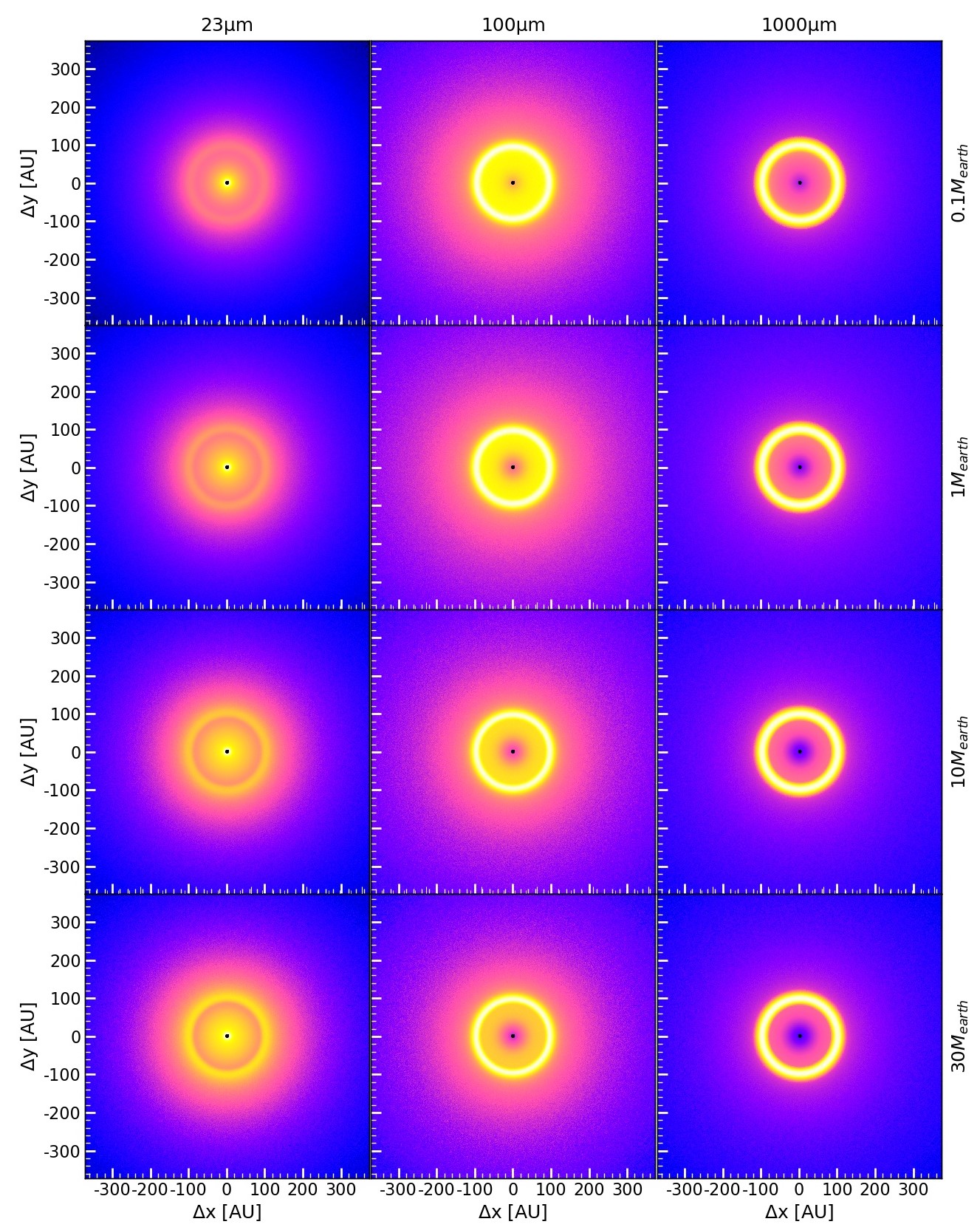}
    \caption{Thermal emission maps for discs around a G-type star at a distance of 100~AU and an age of 1~Gyr. From left to right: observational wavelengths of 23, 100 and 1000$\mum$. From top to bottom: disc masses of 0.1, 1, 10 and 30~$M_\text{earth}$.
    }
     \label{fig:PR-drag_thermalemission}
\end{figure*}

\begin{figure*}
\centering
	\includegraphics[width=\textwidth]{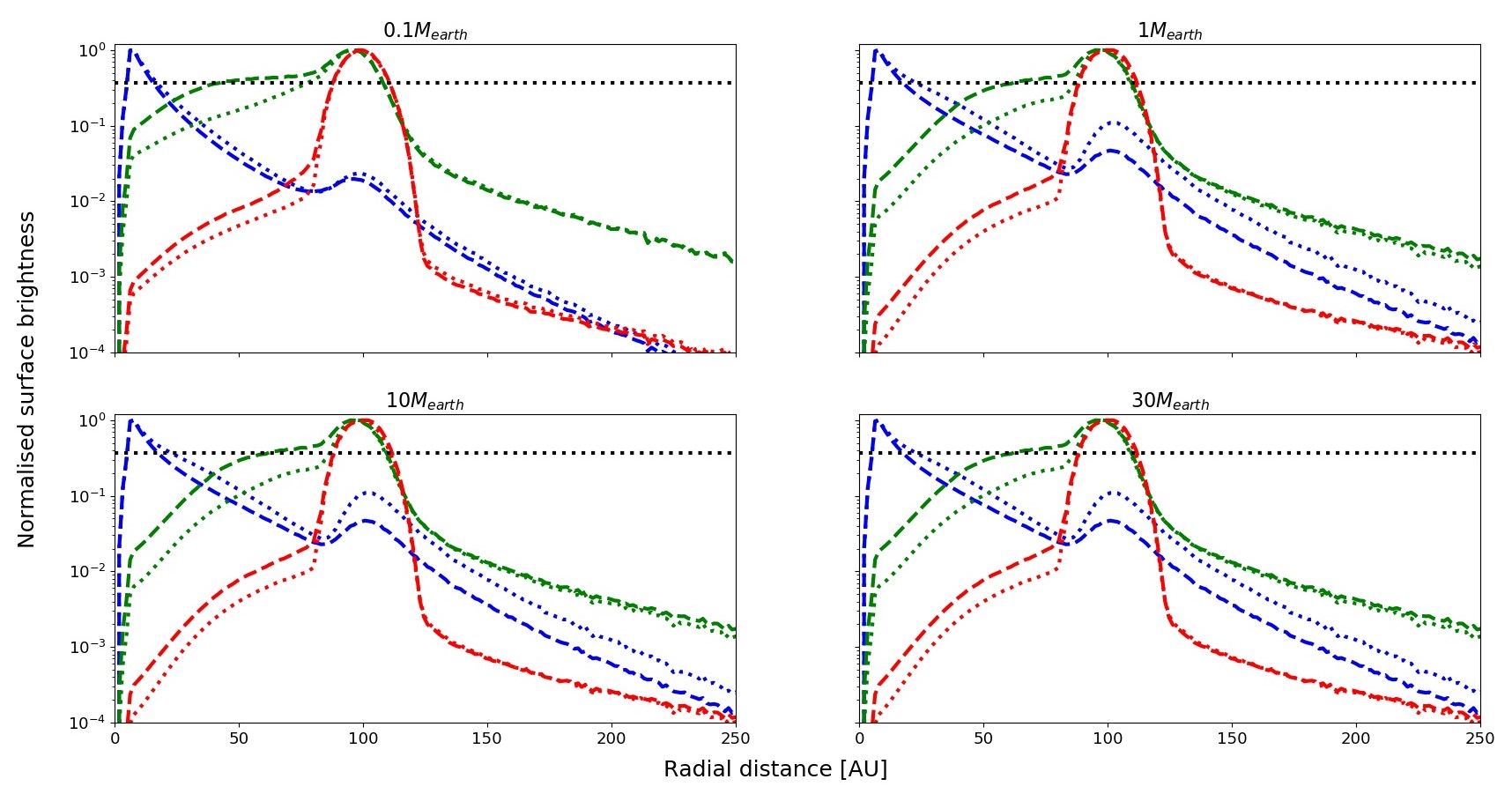}
    \caption{Normalised surface brightness as a function of distance for different disc masses around a G-type star. Profiles are shown for 23$\mum$ (blue), 100$\mum$ (green) and 1000$\mum$ (red). Dotted lines represent an age of 100~Myr, dashed lines an age of 1~Gyr.
    }
     \label{fig:PR-drag_profile}
\end{figure*}

In Figs.~\ref{fig:PR-drag_thermalemission} and \ref{fig:PR-drag_profile} we present the thermal emission maps and surface brightness profiles for the different disc masses at 100~Myr and 1~Gyr. 
It is possible to identify the planetesimal belt location for discs between 0.1 and 30~$M_\text{earth}$ easily at wavelengths of 100 and 1000$\mum$ despite the occurring inward drift.
The situation is more difficult at 23$\mum$ where the inward drift of micron-sized particles leads to an increase in surface brightness close to the star so that the actual planetesimal belt shows only a fraction of this maximum surface brightness (2-10\%). However, the belt location is distinguishable from the position of the surface brightness peak close to the star. In this case coronagraphic observations can help to suppress the peak close to the star.

\subsection{Early-type stars}

\begin{figure*}
\centering
	\includegraphics[width=0.8\textwidth]{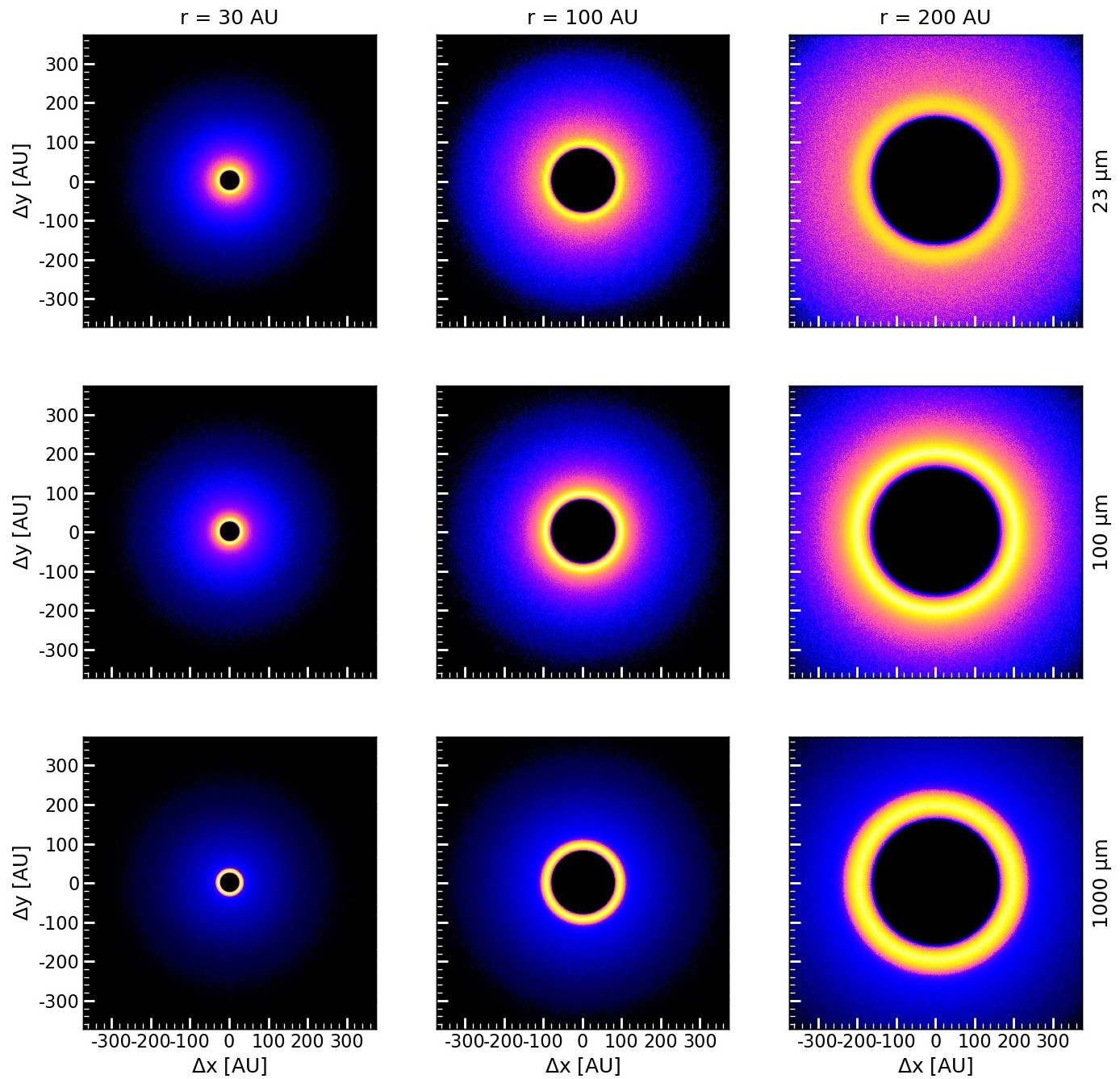}
    \caption{Same as Fig.~\ref{fig:ACE_Mstar}, but for an A-type star with $L/L_\odot=30.0$.}
     \label{fig:ACE_Astar}
\end{figure*}

\begin{figure*}
\centering
	\includegraphics[width=\textwidth]{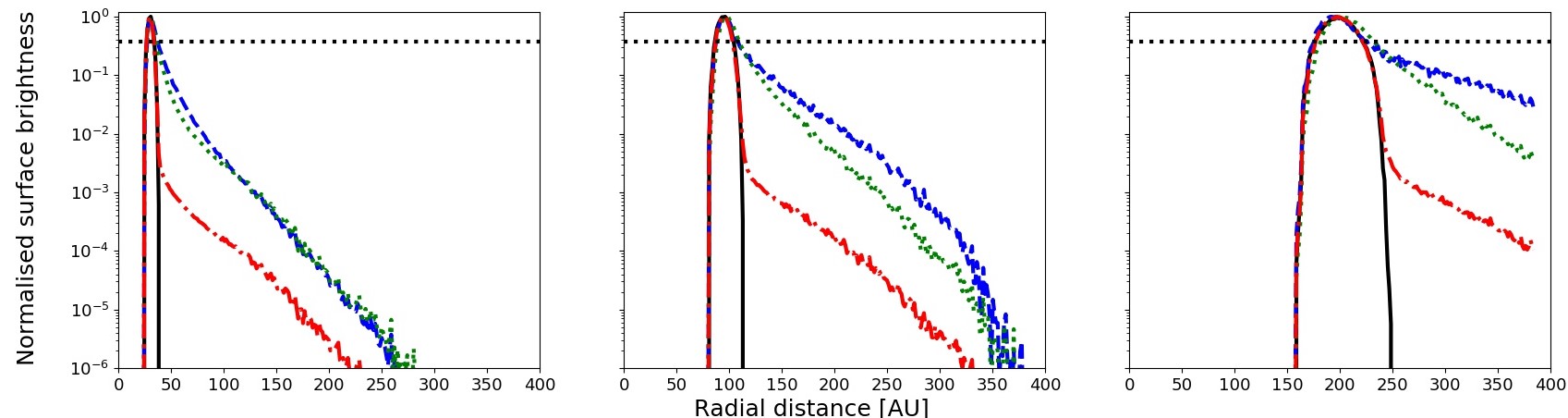}
    \caption{Same as Fig.~\ref{fig:ACE_Mstar_Profile}, but for an A-type star with $L/L_\odot=30.0$.
    }
     \label{fig:ACE_Astar_Profile}
\end{figure*}

In Figs.~\ref{fig:ACE_Astar} and \ref{fig:ACE_Astar_Profile} we show the thermal emission maps and radial profiles for an A-type star with $L/L_\odot=30.0$.
Based on our test results for the G-type stars (see Sect.~4.2.2) and to speed up the simulations, we only present and discuss here the runs without P-R drag.

Similar to the maps of discs around the G-type star we see a broadening of the dust disc due to radiation pressure. Using the location of the $1/e$ value of the surface brightness peak as disc extension we see that the broadening in the far-infrared is comparable to the broadening in the mid-infrared in contrast to the discs around the G-type star where only the mid-infrared images show a subtle broadening. At lower fractions of the surface brightness peak the disc broadening becomes more prominent.
At sub-mm wavelengths, the discs are most confined while the extension increases towards shorter wavelengths.
We would expect this outcome, since we know that smaller grains are more affected by radiation pressure and are better traced at shorter wavelengths. 
Considering the parts of high surface brightness values the profiles are comparable to the profiles of discs around the G-type star, but nevertheless the broadening is more prominent.

\subsection{Additional model setups}

We extend our results from the previous sections by investigating additional model setups, now focusing on discs with a radius of $100$~AU around a solar-type star. 
We test a broader planetesimal belt and consider a higher disc excitation level. 

\subsubsection{With a broad ring}

In our previous models we assumed a narrow planetesimal belt which was justified by earlier studies showing that the majority of the discs is confined into narrow rings \citep[e.g.,][]{kennedy-wyatt-2010}.
Nevertheless, recent studies showed that discs are also found with extended planetesimal belts \citep[e.g.,][]{matra-et-al-2018}, and  hence we want to include a broad birth ring into our analysis of the peak emission of the surface brightness.

\begin{figure*}
\centering
	\includegraphics[width=0.9\textwidth]{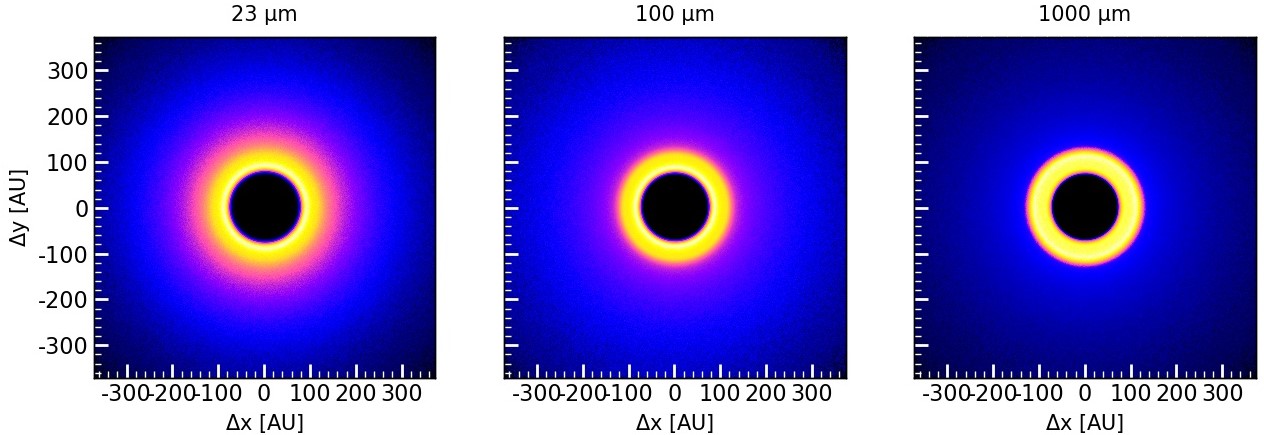}
    \caption{Thermal emission maps assuming a G-type central star and a parent belt centred at 100~AU, for a relative belt width of $\pm$20$\%$. From left to right: the emission at 23, 100 and 1000~$\mum$.}
     \label{fig:ACE_Gstar_Width}
\end{figure*}

\begin{figure}
\centering
	\includegraphics[width=0.35\textwidth, angle=-90]{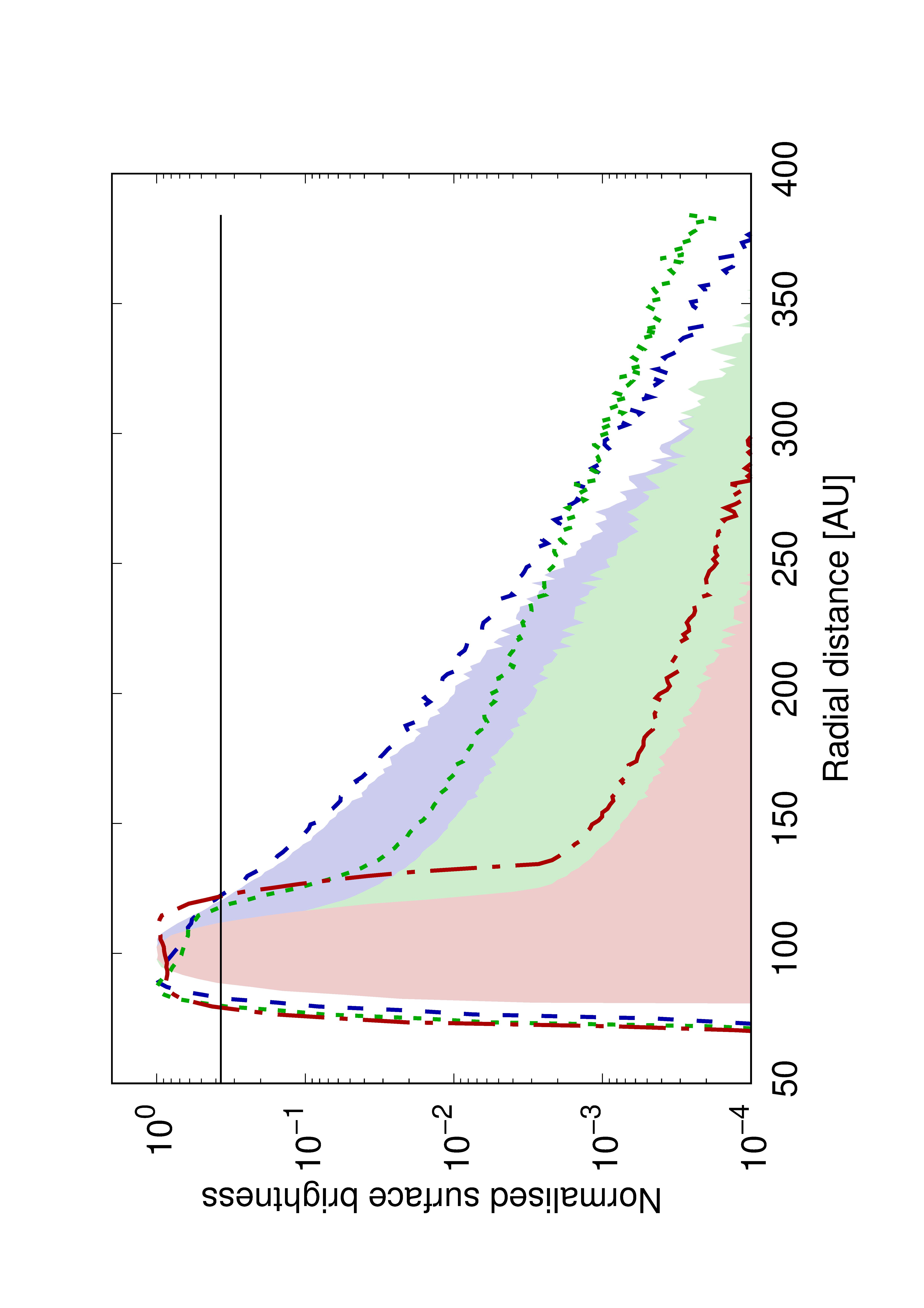}
    \caption{Surface brightness profiles assuming a G-type central star and a parent belt centred at 100~AU. Blue shows the profile at 23~$\mum$, green at 100~$\mum$ and red at 1000~$\mum$. The filled areas represent the reference case with a belt width of $\pm10\%$. The lines correspond to a width of $\pm20\%$. The black solid line shows the $1/e$ level of the maximum.}
     \label{fig:ACE_Gstar_Width_Profile}
\end{figure}

While for the former disc we assume a belt width of $100\pm10$~AU we now use a planetesimal ring of $100\pm20$~AU. 
The thermal emission maps at 23, 100 and 1000$\mum$ are presented in Fig.~\ref{fig:ACE_Gstar_Width}. The radial profiles are shown in Fig.~\ref{fig:ACE_Gstar_Width_Profile}.

Similar to the default case with a 10\% width of the belt, we see a broadened surface brightness profile caused by radiation pressure. The dust disc is most confined at 1000$\mum$ while it shows a broad extension towards shorter wavelengths.
The resulting dust disc of a planetesimal ring with a width of 20\% is more extended in comparison to the disc stemming from a parent belt with a width of 10\%. 
Its surface brightness profile shows a flatter decrease with increasing radial distance.
For example, if we assume the level of the normalised surface brightness to be $10^{-4}$, the radial extension at 1000$\mum$ is $\sim$250~AU considering a belt width of 10\% and 300~AU for a width of 20\%. 

The maximum of the surface brightness is distributed over a larger distance range, hence we see a plateau in the radial profile for the case of the broad birth ring. However, this plateau is symmetrically extended around 100~AU, so that we are still able to define the radius of the disc as the peak in the surface brightness by inferring the centre of the plateau.

\subsubsection{With increased stirring level}

\begin{figure*}
\centering
	\includegraphics[width=0.9\textwidth]{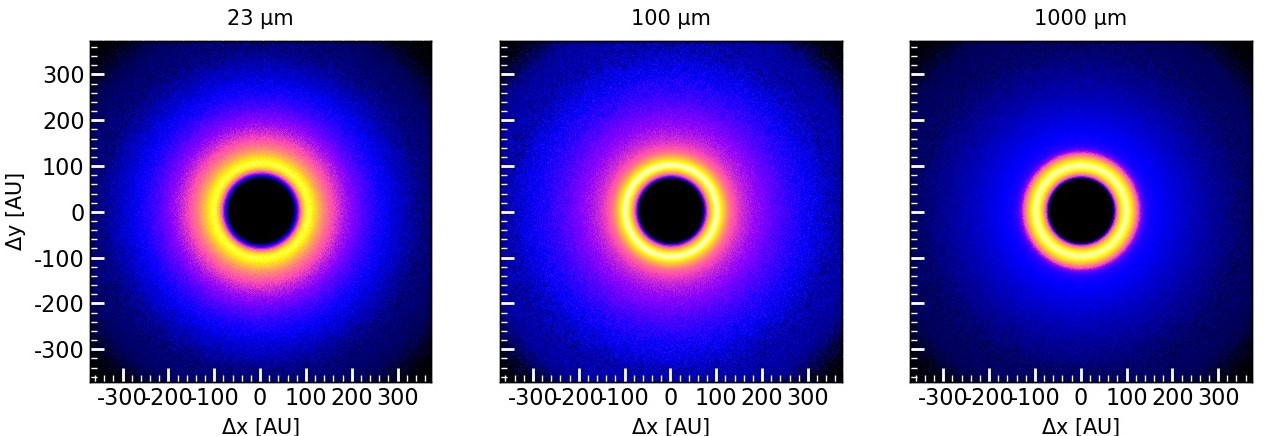}
    \caption{Same as Fig.~\ref{fig:ACE_Gstar_Width}, 
    but for the average eccentricity of the planetesimals of 0.2.}
     \label{fig:ACE_Gstar_Excitation}
\end{figure*}

\begin{figure}
\centering
	\includegraphics[width=0.35\textwidth, angle=-90]{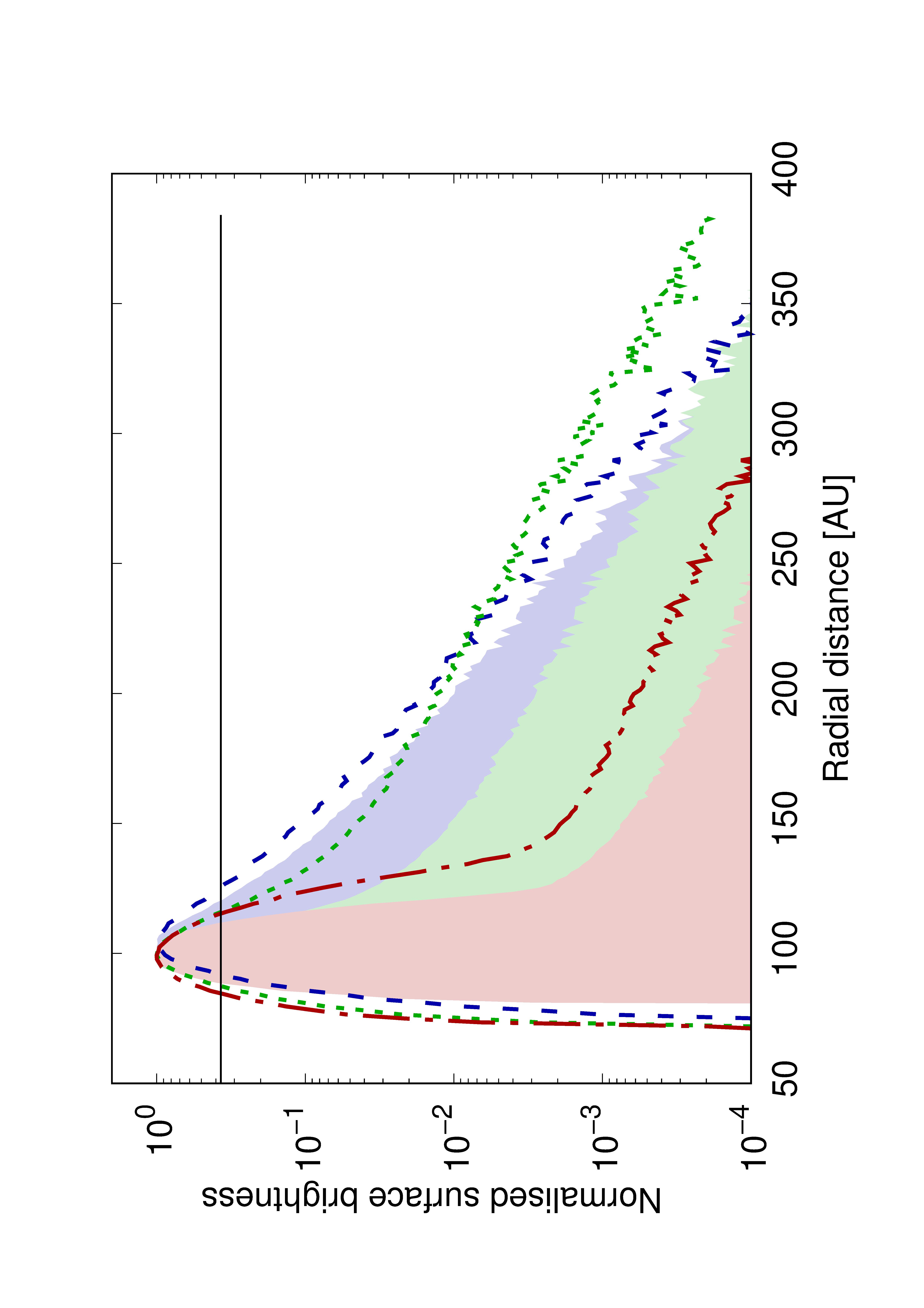}
    \caption{Surface brightness profiles assuming a G-type central star and a parent belt centred at 100~AU. Blue shows the profile at 23~$\mum$, green at 100~$\mum$ and red at 1000~$\mum$. The filled areas represent an average eccentricity of 0.1 for the planetesimals while the lines show the result for a value of 0.2. The black solid line shows the $1/e$ level of the maximum.}
     \label{fig:ACE_Gstar_Excitation_Profile}
\end{figure}

Another parameter to vary is the stirring level of the parent belt.
The Edgeworth-Kuiper belt of our Solar system possesses an average planetesimal eccentricity of 0.08 \citep{vitense-et-al-2012} which justifies our assumption of 0.1 for our models. 
However, higher stirring levels are possible, especially around earlier-type stars \citep{pawellek-krivov-2015}.
As we did before, we use the 
parent belt at 100~AU with the G-type host star and increase
the average eccentricity of the parent bodies to 0.2 in contrast to the default case of 0.1. 
The results are shown in Figs.~\ref{fig:ACE_Gstar_Excitation} and \ref{fig:ACE_Gstar_Excitation_Profile}.

The higher excitation of the parent bodies leads to a broadening of the disc similar to the assumption of a broader planetesimal belt. Thus, the resulting thermal emission maps and surface brightness profiles look similar compared to the results of the last section and reveal a degeneracy between both scenarios. 
However, by looking at the shape of the surface brightness peak we can see that it is still confined around the birth ring location and not extended over a larger distance range (see Figs.\ref{fig:ACE_Gstar_Width_Profile} and \ref{fig:ACE_Gstar_Excitation_Profile}). 
Thus, if we are able to analyse the shape of the surface brightness peak, we can determine whether the disc extension is caused by a higher disc excitation or a broader planetesimal belt. Nevertheless, with current instrument facilities and observational issues (see Section~\ref{sec:convolution}) this analysis remains difficult, so that we are hardly able to break the aforementioned degeneracy. 

We also note that the radially extended belts, such as those listed in \citet{matra-et-al-2018}, can be viably attributed to the ``scattered discs'', similar to the scattered disc in our solar system \citep{wyatt-et-al-2017,geiler-et-al-2019}. Such a scattered disc could be composed of planetesimals in orbits with a narrow range of semimajor axes, but large eccentricities. For instance, the values in the range $e \sim 0.3$--$0.5$ have been found to best reproduce the scattered disc of HR~8799 \citep{geiler-et-al-2019}. Thus the run presented here can also be considered as a proxy of what could be expected for a ``moderately scattered'' disc. Since the result does not change much compared to the reference setup and taking into account that running ACE for higher eccentricities would be very expensive computationally, we do not present here a model for eccentricities above 0.2.

\subsection{Simulating realistic observations}
\label{sec:convolution}

To get an impression of  the observable disc radii, we simulate realistic observations for \citep[MIRI/JWST, ][]{rieke-et-al-2015} and for ALMA. To do that, the distance of our disc models with the G-type central star is set to 50~pc.

\subsubsection{JWST observations}

\begin{figure*}
\centering
	\includegraphics[width=0.85\textwidth]{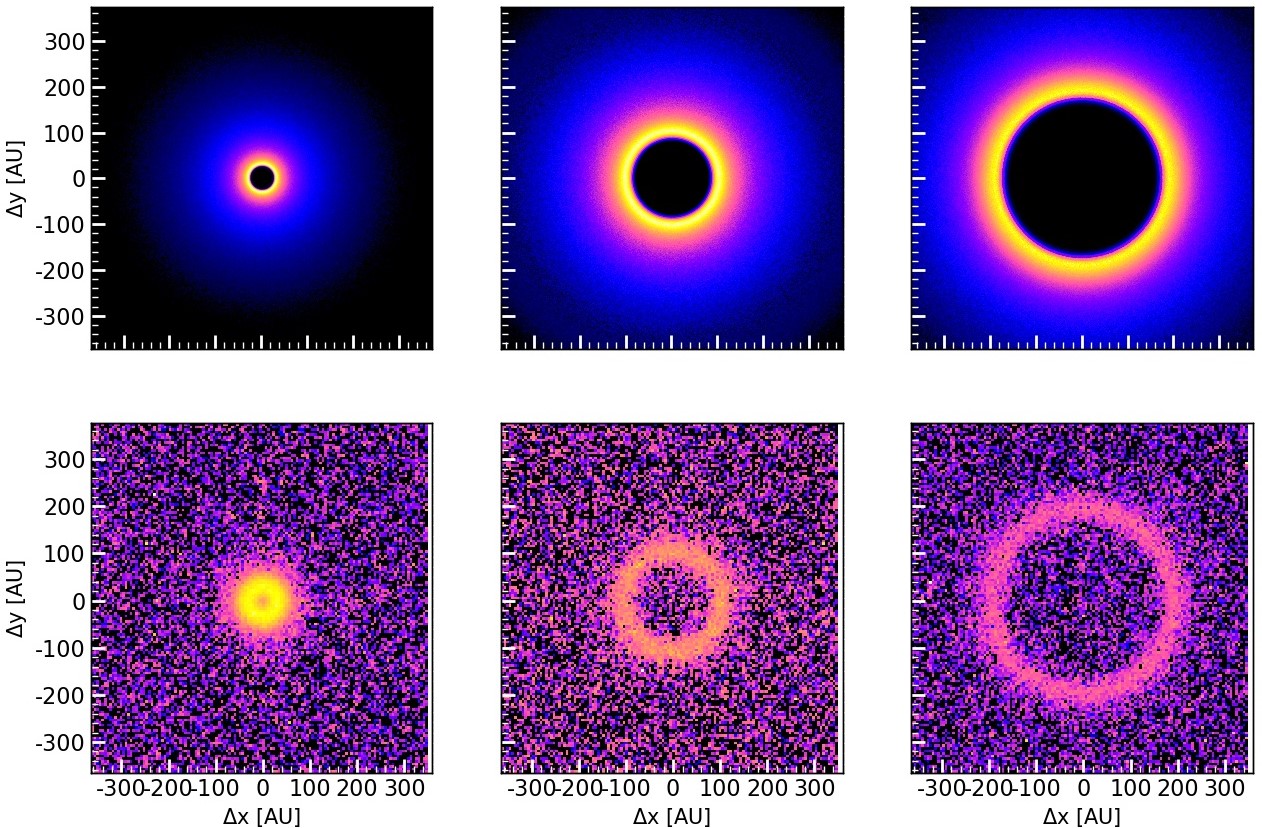}
    \caption{Thermal emission maps for a G-type star at 23~$\mum$. Upper panels: Model discs at 30, 100 and 200~AU. Lower panels: Model discs convolved with the PSF of MIRI/JWST.}
     \label{fig:ACE_Gstar_Convolution_JWST}
\end{figure*}

For JWST, we use the MIRISim tool\footnote{http://miri.ster.kuleuven.be/bin/view/Public/\\MIRISim\_Public\#MIRISim} provided by the MIRI European Consortium and specifically designed to simulate MIRI observations.
The resulting raw data from the simulator are treated with the preliminary JWST pipeline \citep{bushouse-et-al-2017}. We are aware that the results shown below might change with future versions of the simulator and the pipeline. 
We apply the imager mode and assume the F2100W wavelength filter with a band width of 5$\mum$. For the observational setup, $10$ integrations with $30$ frames each are used, leading to an exposure time of $\sim13$ minutes. 
The results are depicted in Fig,~\ref{fig:ACE_Gstar_Convolution_JWST}.

The discs get fainter with increasing disc radius. This is no surprise since the mid-infrared is more sensitive to warm discs compared to cold ones. However, we are able to resolve all of them and thus, MIRI will be useful for observations of debris discs close to the star. 
The discs show a slight broadening compared to the model images, but the position of the surface brightness peak stays the same. 
We do not see the broadening of the discs as shown in the models, but this might change with an increase of the exposure time.

\subsubsection{ALMA observations}

We use the ALMA Observation Support Tool\footnote{\url{http://almaost.jb.man.ac.uk/}} 
(Version 6.0) for the millimeter continuum observations.  We adopted the ALMA Cycle\,6 
C43-5 configuration that provides baselines ranging from 15 to 1400\,m. The target was 
placed at a declination of $-$23 degrees. Using an integration time of 6 hours and 
assuming natural weighting we inferred a beam size of 0.265{\arcsec}$\times$0.226{\arcsec}, 
and a beam position angle (PA) of 86\fdg1.

\begin{figure*}
\centering
	\includegraphics[width=0.85\textwidth]{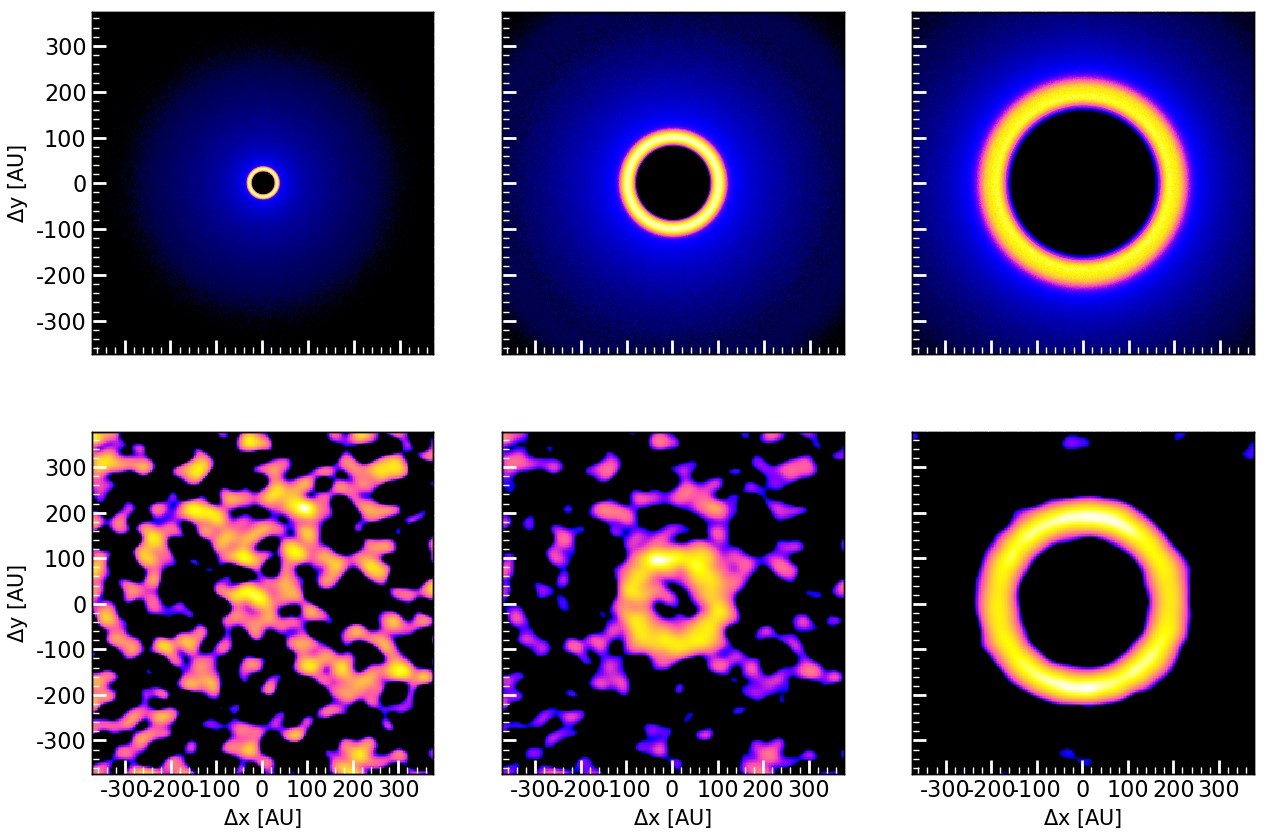}
    \caption{Thermal emission maps for a G-type star at 1000~$\mum$. Upper panels: Model discs at 30, 100 and 200~AU. 
    Lower panels: Simulated ALMA images compiled by the ALMA Observation Support Tool.}
     \label{fig:ACE_Gstar_Convolution_ALMA}
\end{figure*}

Assuming a bandwidth of 7.5~GHz and a precipitable water vapour column of 
1.8~mm this setup offers us a 1$\sigma$ sensitivity of 8.2~$\mu$Jy/beam.
The obtained simulated images are shown in Fig.~\ref{fig:ACE_Gstar_Convolution_ALMA}.

We are able to resolve the more massive discs at 100 at 200~AU, but cannot detect the lowest mass disc at 30~AU which is simply too faint at millimetre wavelengths for ALMA.
In the former two images the position of the peak emission traces well
the parent planetesimal belts.

\section{Conclusions}
\label{sec:conclusions}

In this paper, we analysed the possibility of locating parent planetesimal belts of debris discs by observing their dust emission at different wavelengths. 
For this purpose we generated models of fiducial discs of different radii around main-sequence stars of various spectral types covering a broad luminosity range. We used the ACE code \citep{loehne-et-al-2017} to follow the collisional evolution of discs
and then created their thermal emission images and surface brightness profiles at different wavelengths from mm down to mid-infrared.

For the reference model setup (excitation level of 0.1, a relative belt width of $10\%$, without Poynting-Robertson and stellar wind drag), we found that it is possible to trace the planetesimal belts of all radii tested (from $30$~AU to $200$~AU) around all host stars (from A to M) and at all wavelengths investigated (from $1$~mm down to $23 \mum$). Compared to the extension of the birth ring, the discs show a broadening in their surface brightness due to radiation pressure. This broadening is stronger towards earlier-type stars. However, the peak of the surface brightness stays at nearly the same location for all discs, independent of the spectral type of the host star. We have also checked that this conclusion remains valid if Poynting-Robertson drag is included, as long as the peak normal optical depth of the disc is above $\sim 10^{-5}$.

However, for M-stars we expect stronger stellar winds compared to earlier-type stars. By including a moderately strong stellar wind into our models we found that the position of the surface brightness peak wanders inwards even for long wavelengths. 
This will underestimate the radius of the birth ring.
The estimation error will be larger for more compact discs and at shorter wavelengths. 

To test the robustness of our results, we varied the relative width of the planetesimal belt and the assumed stirring level.
For an increased belt width, 
the surface brightness profile develops an extended plateau. 
The higher stirring level leads to a comparable disc broadening, yet the surface brightness peak stays more distinct. Thus, by analysing the shape of the surface brightness profile it might be possible to differentiate between the effects of a broader planetesimal belt and a higher stirring level.

To get a more realistic picture of what can be expected from observations with specific instruments, we simulated images for MIRI/JWST and ALMA assuming a typical distance of 50~pc. 
We found that both instruments are able to resolve the discs considered in this study and that the position of the surface brightness peak stays at the location of the planetesimal belt. 
While ALMA is most efficient at tracing large debris discs, MIRI will be more suitable to locate the warmer belts close to the star. 

In summary, we conclude that 
tracing the planetesimal belts of debris discs around early- and solar-type stars is possible even at mid-infrared wavelengths. Observations at (sub-)mm wavelengths are not really necessary. For M-type stars, however, such observations are indispensable.

\section*{Acknowledgements}
We are grateful to the anonymous referee for their detailed comments that greatly helped to improve the manuscript.
NP is thankful to Torsten L\"ohne for useful discussions of the collisional modelling. NP thanks J\"urgen Schreiber and Jeroen Bouwman for sharing their knowledge of MIRI and its simulator.
AM acknowledges the support of the Hungarian National Research, De-
velopment  and  Innovation  Office  NKFIH  Grant  KH-130526. 
AVK acknowledges support by the Deutsche Forschungsgemeinschaft (DFG), grants No. Kr~2164/13-1 and Kr~2164/15-1. 
This material is based upon work supported by the National Aeronautics and Space Administration under Agreement No. NNX15AD94G for the program “Earths in Other Solar Systems”. The results reported herein benefited from collaborations and/or information exchange within NASA’s Nexus for Exoplanet System Science (NExSS) research coordination network sponsored by NASA’s Science Mission Directorate. 




\newcommand{\AAp}      {A\& A}
\newcommand{\AApR}     {Astron. Astrophys. Rev.}
\newcommand{\AApS}     {AApS}
\newcommand{\AApSS}    {AApSS}
\newcommand{\AApT}     {Astron. Astrophys. Trans.}
\newcommand{\AdvSR}    {Adv. Space Res.}
\newcommand{\AJ}       {AJ}
\newcommand{\AN}       {AN}
\newcommand{\AO}       {App. Optics}
\newcommand{\ApJ}      {ApJ}
\newcommand{\ApJL}     {ApJL}
\newcommand{\ApJS}     {ApJS}
\newcommand{\ApSS}     {Astrophys. Space Sci.}
\newcommand{\ARAA}     {ARA\& A}
\newcommand{\ARevEPS}  {Ann. Rev. Earth Planet. Sci.}
\newcommand{\BAAS}     {BAAS}
\newcommand{\CelMech}  {Celest. Mech. Dynam. Astron.}
\newcommand{\EMP}      {Earth, Moon and Planets}
\newcommand{\EPS}      {Earth, Planets and Space}
\newcommand{\GRL}      {Geophys. Res. Lett.}
\newcommand{\JGR}      {J. Geophys. Res.}
\newcommand{\JOSAA}    {J. Opt. Soc. Am. A}
\newcommand{\MemSAI}   {Mem. Societa Astronomica Italiana}
\newcommand{\MNRAS}    {MNRAS}
\newcommand{\PASJ}     {PASJ}
\newcommand{\PASP}     {PASP}
\newcommand{\PSS}      {Planet. Space Sci.}
\newcommand{\prospie}  {Proceedings of the SPIE}
\newcommand{\RAA}      {Research in Astron. Astrophys.}
\newcommand{\SolPhys}  {Sol. Phys.}
\newcommand{\SolSysRes}{Sol. Sys. Res.}
\newcommand{\SSR}      {Space Sci. Rev.}

\input{Bib.bbl.std}






\bsp	
\label{lastpage}
\end{document}